\documentclass[twocolumn]{aastex61}

\renewcommand{\thefootnote}{\fnsymbol{footnote}}
\let\OldS\S
\renewcommand{\S}{\OldS{}}

\received{April, 2019}
\revised{June, 2019}
\accepted{June, 2019}
\submitjournal{ApJ}

\shorttitle{Ly$\alpha$ nebula around J0952+0114}
\shortauthors{Marino et al.}

\begin{document}

\title{A giant Ly$\alpha$ nebula and a small-scale clumpy outflow in the system of the exotic quasar J0952+0114 unveiled by MUSE\footnote{Based on observations obtained at the Very Large Telescope (VLT) of the European Southern Observatory, Paranal, Chile (ESO Programme ID 096.A-0345).}.}

\correspondingauthor{Raffaella Anna Marino}
\email{marinor@phys.ethz.ch}

\author[0000-0002-8559-6565]{Raffaella Anna Marino}
\affil{Department of Physics, ETH Z$\ddot{u}$rich,Wolfgang-Pauli-Strasse\,27, 8093\,Z$\ddot{u}$rich, Switzerland}

\author{Sebastiano Cantalupo}
\affiliation{Department of Physics, ETH Z$\ddot{u}$rich,Wolfgang-Pauli-Strasse\,27, 8093\,Z$\ddot{u}$rich, Switzerland}

\author{Gabriele Pezzulli}
\affiliation{Department of Physics, ETH Z$\ddot{u}$rich,Wolfgang-Pauli-Strasse\,27, 8093\,Z$\ddot{u}$rich, Switzerland}

\author{Simon J$.$ Lilly}
\affiliation{Department of Physics, ETH Z$\ddot{u}$rich,Wolfgang-Pauli-Strasse\,27, 8093\,Z$\ddot{u}$rich, Switzerland}

\author{Sofia Gallego}
\affiliation{Department of Physics, ETH Z$\ddot{u}$rich,Wolfgang-Pauli-Strasse\,27, 8093\,Z$\ddot{u}$rich, Switzerland}

\author{Ruari Mackenzie}
\affiliation{Department of Physics, ETH Z$\ddot{u}$rich,Wolfgang-Pauli-Strasse\,27, 8093\,Z$\ddot{u}$rich, Switzerland}

\author{Jorryt Matthee}
\affiliation{Department of Physics, ETH Z$\ddot{u}$rich,Wolfgang-Pauli-Strasse\,27, 8093\,Z$\ddot{u}$rich, Switzerland}

\author{Jarle Brinchmann}
\affiliation{Instituto de Astrof{\'\i}sica e Ci{\^e}ncias do Espaço, Universidade do Porto, CAUP, Rua das Estrelas, PT4150-762 Porto, Portugal}
\affiliation{Leiden Observatory, Leiden University, P.O. Box 9513, 2300 RA, Leiden, The Netherlands}

\author{Nicolas Bouch\'e}
\affiliation{Institut de Recherche en Astrophysique et Plan\'etologie (IRAP), Universit\'e de Toulouse, CNRS, UPS, F-31400 Toulouse, France}
\affiliation{Univ Lyon1, Ens de Lyon, CNRS, Centre de Recherche Astrophysique de Lyon UMR5574, F-69230 Saint-Genis-Laval, France}

\author{Anna Feltre}
\affiliation{Univ Lyon1, Ens de Lyon, CNRS, Centre de Recherche Astrophysique de Lyon UMR5574, F-69230 Saint-Genis-Laval, France}
\affiliation{Scuola Internazionale Superiore di Studi Avanzati (SISSA), Via Bonomea 265, I-34136, Trieste, Italy}

\author{Sowgat Muzahid}
\affiliation{Leiden Observatory, Leiden University, PO Box 9513, NL-2300 RA Leiden, the Netherlands}

\author{Ilane Schroetter}
\affiliation{GEPI, Observatoire de Paris, PSL Universit\'e, CNRS, 5 Place Jules Janssen, 92190 Meudon, France}


\author{Sean D. Johnson}
\affiliation{Department of Astrophysical Science, 4 Ivy Lane, Princeton University, Princeton, NJ 08644, USA}
\affiliation{The Observatories of the Carnegie Institution for Science, 813 Santa Barbara Street, Pasadena, CA 91101, USA}

\author{Themiya Nanayakkara}
\affiliation{Leiden Observatory, Leiden University, PO Box 9513, NL-2300 RA Leiden, the Netherlands}
%



\begin{abstract}

The well-known quasar SDSS J095253.83+011421.9 (J0952+0114) at \textit{z}\,$=$\,3.02 has one of the most peculiar spectra discovered so far, showing the presence of narrow Ly$\alpha$ and broad metal emission lines. Although recent studies have suggested that a Proximate Damped Ly$\alpha$ system (PDLA) causes this peculiar spectrum, the origin of the gas associated with the PDLA is unknown.  Here we report the results of MUSE observations that reveal a new giant ($\approx$ 100 physical kpc) Lyman $\alpha$ nebula. The detailed analysis of the Ly$\alpha$ velocity, velocity dispersion, and surface brightness profiles suggests that the J0952+0114 Ly$\alpha$ nebula shares similar properties of other QSO nebulae previously detected with MUSE, implying that the PDLA in J0952+0144 is covering only a small fraction of the QSO emission solid angle.  We also detected bright and spectrally narrow \ion{C}{4}\,$\lambda$1550 and \ion{He}{2}\,$\lambda$1640 extended emission around J0952+0114 with velocity centroids similar to the peak of the extended and central narrow Ly$\alpha$ emission. The presence of a peculiarly bright, unresolved, and relatively broad \ion{He}{2}\,$\lambda$1640 emission in the central region at exactly the same PDLA redshift hints at the possibility that the PDLA originates in a clumpy outflow with a bulk velocity of about 500 km\,s$^{-1}$. The smaller velocity dispersion of the large scale Ly$\alpha$ emission suggests that the high-speed outflow is confined to the central region. Lastly, the derived spatially resolved \ion{He}{2}/Ly$\alpha$ and \ion{C}{4}/Ly$\alpha$ maps show a positive gradient with the distance to the QSO hinting at a non-homogeneous ionization parameter distribution.\\

\end{abstract}
\keywords{intergalactic medium - quasars: general - quasars: emission lines - quasars: individual(SDSS J095253.83+011421.9) - techniques: imaging spectroscopy}



\def\nodata{ ~$\cdots$~}
\section{Introduction} \label{sec:intro}

\citet[][henceforth H04]{2004AJ....128..534H} published the discovery of an exotic  $z$\,$=$\,3.02 quasar (QSO), SDSS J095253.83+\,  011421.9 (hereafter J0952+0114) lacking a broad Ly$\alpha$ emission line. 

\begin{figure*}[ht!]
\includegraphics[angle=270,scale=0.6]{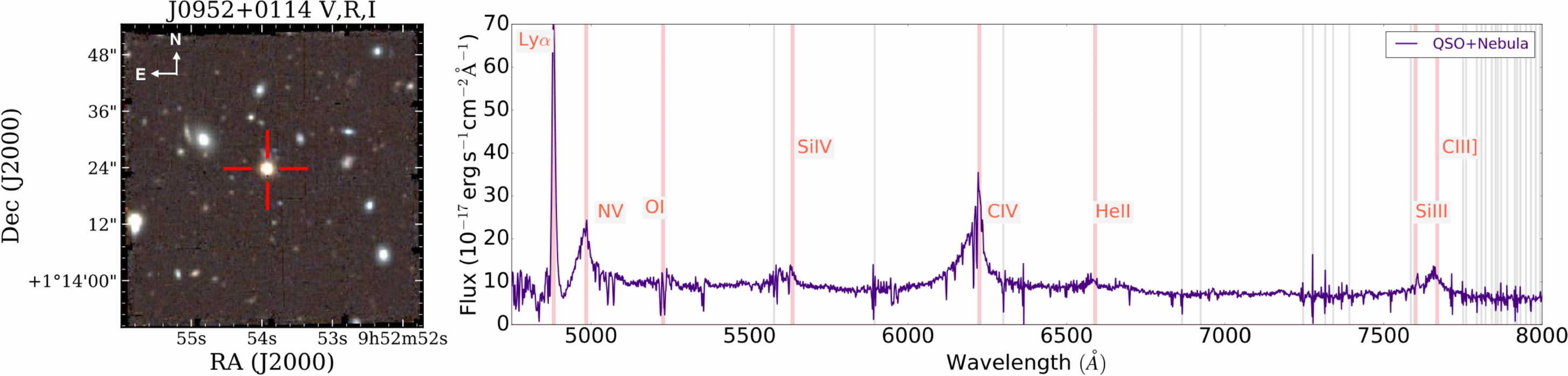}
\caption{ The combined MUSE RGB image and optical spectrum of the J0952+0114 system. \textit{Left Panel:} The RGB colors correspond to the pseudo broad V$-$, R$-$, and I$-$band images obtained from the MUSE datacubes. The QSO location is marked by the red cross. North is up and east is to the left. \textit{Right Panel:} J0952+0114 MUSE 1D optical spectrum of the central region extracted within the 3D Ly$\alpha$ segmentation mask of the nebula (i$.$ e$.$ includes the contributions of both QSO and nebula, see Section\,\ref{sec:analysis} for more details). The expected emission lines are labelled in orange and the grey vertical lines indicate the position of the most prominent skylines.}\label{fig:RGB_spectrum}
\end{figure*}

H04 invoked several models to explain the observed emission line properties, including dust extinction in the Broad Line Region (BLR); anisotropic Ly$\alpha$ emission; unusual physical conditions of the BLR (high ionization parameters and a peculiar configuration of the emitting clouds); intrinsic moderate absorption by the resonant transition of \ion{N}{5} (in the red wing) and Ly$\alpha$ (in the blue one) to the total Ly$\alpha$ emission (for a detailed discussion see Section 3 in H04). Finally, they conclude that the dominant effect responsible of the broad Ly$\alpha$ weakness is the presence of very dense gas (n$_{H}\,=$10$^{15}$ cm$^{-3}$) that is suppressing the broad line component as well as increasing the collisionally excited metal-line emission. 

The partially unsatisfactory interpretation of such peculiar spectral properties has intrigued the community for more than a decade and, therefore, motivated a BOSS \citep[Baryon Oscillation Spectroscopic Survey][]{2013AJ....145...10D} follow up by \citet[][henceforth J16]{2016ApJ...821....1J} to better determine the true line profiles (and possible physical reasons) behind such a unique QSO spectrum.  

Taking advantage of both the BOSS wider wavelength coverage (3600$\mathrm{\AA}$\,-\,10400$\mathrm{\AA}$) and higher resolution (R$\sim$2000) with respect to the SDSS-I/II \citep{2000AJ....120.1579Y} data available, J16 detected the clear imprints of a high column density Proximate Damped Lyman $\alpha$ Absorption (PDLA) system at $\textit{z}_{\mathrm{abs}}=$ 3.01 toward the line-of-sight of this QSO (from the damped Ly$\beta$ absorption, the high-order Lyman-series absorptions, the Lyman limit absorption edge and metal-absorption lines). As described in J16, this PDLA is responsible for absorbing the intrinsic Ly$\alpha$ emission of the QSO, while the observed residual narrow Ly$\alpha$ emission line (FWHM\,$\sim$\,1000 km\,s$^{-1}$)  arises from gas distributed over a larger area than the cross-section size of the PDLA cloud (and possibly more extended than the Narrow Line Region, NLR). 
Unfortunately, the Ly$\alpha$ emission was not spatially resolved in their SDSS data and solely based on the large line velocity width and covering factor, J16 suggested that such Ly$\alpha$ emission could originate from outflows driven by the QSO.

In addition, J16 found that this system is not so rare, which was also confirmed by \citet{2013A&A...558A.111F} and by the recent works of \citet{2016MNRAS.461.1816F, 2018MNRAS.477.5625F} which identified $\approx$ 400 eclipsing DLAs in the SDSS-III DR12. In particular, \citet{2018MNRAS.477.5625F} distinguished three populations of PDLAs based on the narrow Ly$\alpha$ strength. Their interpretation was that PDLAs result from the interaction between infalling and outflowing gas at different distances from the QSO, supporting the outflow scenario proposed by J16 for the J0952+0114 system.


Certainly, as suggested by previous studies, direct observations of extended Ly$\alpha$ emission (the ``seeing fuzz" supposed by J16)  and detailed kinematic analysis are necessary to test the outflow hypothesize and develop a better understanding of the QSO J0952+0114 and its environment.

In this paper, we will make these steps forwards and present the Integral Field Unit (IFU) follow-up observations of J0952+0114 with VLT/MUSE, which have the spatial and spectral domains necessary to provide new insights into the neighborhood of this exotic QSO. The key questions that drive our investigation are: where does the observed narrow Ly$\alpha$ emission come from? How different (or similar) is this QSO with respect to the other MUSE QSO snapshot (1 hour) fields? Is there any on-going outflow event in the J0952+0114 system?
 
We report the detection of extended emission for several lines, including Ly$\alpha$, \ion{C}{4} and \ion{He}{2}, 
and we investigate possible explanations for the missing broad Ly$\alpha$ in J0952+0114 and the kinematics of its surrounding gas. Finally, we interpret the possible physical configuration of the different components that are part of the J0952+0114 system (see Fig.\,\ref{fig:cartoon}).

The paper is organized as follows. In \S{} 2 we describe the MUSE observations, data reduction and post-processing. In \S{} 3 we detail the analysis performed on the J0952+0114 datacube. In \S{} 4 we present our results and in \S{} 5 we discuss our findings. Finally, we summarize our conclusions in \S{} 6.

We adopt a flat $\Lambda$CDM cosmology with Wilkinson Microwave Anisotropy Probe 9 cosmological parameters of $\Omega_{\Lambda}$\,$=$\,0.714, $\Omega_{M}$\,$=$\,0.286 and h\,$=$\,0.693 \citep{2013ApJS..208...19H}, corresponding to $\sim$\, 7.6 kpc/$\textrm{\arcsec}$ at redshift $\sim$\, 3 throughout this work. All wavelengths are specified in vacuum and all magnitudes in the AB system \citep{1983ApJ...266..713O} unless otherwise stated. \\

\section{Observations and Data Reduction} \label{sec:obs}

The QSO J0952+0114 has an estimated systemic redshift of \textit{z}\,$\approx$\,3.020$\,\pm\,$0.005 in the literature obtained from several narrow emission lines detected in both the SDSS and BOSS (H04, J16) spectra. 
Indeed, J0952+0114 exhibits unusual spectral properties with both broad and narrow metal-line emissions but only narrow Ly$\alpha$ emission, as also shown by the MUSE (Multi Unit Spectroscopic Explorer) integrated spectrum in Fig.\,\ref{fig:RGB_spectrum}. 

With its I magnitude of 18.95 \citep{2003AJ....126.2579S}, the quasar is classified as a radio-quiet QSO by \citet{2002AJ....124.2364I} because of its radio-loudness parameter R$_{i}\,<\,$1.024.

Our MUSE observations of J0952+0114 were carried out in March 2016 as part of our Guaranteed Time Observation (GTO) program, in Wide Field  Mode (WFM). MUSE \citep{2010SPIE.7735E..08B} is the Integral Field Spectrograph (IFS) mounted on UT4 at the Very Large Telescope (VLT) in Paranal, Chile. MUSE combines a relatively large field-of-view (FoV in WFM $\approx$ 1\,$\arcmin \times$\,1\,$\arcmin$), an excellent spatial sampling (0.2$\textrm{\arcsec}\,\times$ 0.2$\textrm{\arcsec}$) and spectral resolution (R from $\sim$\,1750 to $\sim$\,3500) from 4750\,$\mathrm{\AA}$ to 9300\,$\mathrm{\AA}$.

We observed the source at the position of $\alpha_{(J2000)}=\,$09\,$:$\,52\,$:$\,53.8 and $\delta_{(J2000)}=\,$+01\,$:$\,14\,$:$\,22 for a total integration time of 1 hour under photometric conditions and a Point Spread Function (PSF) with Full Width Half Maximum (FWHM) of $\sim$ 0.7$\mathrm{\arcsec}$ measured at 7000\,$\mathrm{\AA}$.
The QSO J0952+0114 observations were distributed in four exposures of 15 minutes each, with a dithering pattern smaller than 1$\textrm{\arcsec}$ and a rotation scheme of 90$^{\circ}$ for each individual exposure (see also \citealt[][B16 and M18 hereafter,]{2016ApJ...831...39B, 2018ApJ...859...53M} for more details on the data acquisition strategy).

The reduction of the J0952+0114 data comprises a combination of recipes from the standard ESO MUSE Data Reduction Software \citep[DRS, pipeline version 1.6,][]{2015scop.confE..53W} and from {\ttfamily CubExtractor} software package ({\ttfamily CubEx} in brief, Cantalupo, in prep.; the reader is referred to B16, M18 \citealt{2019MNRAS.482.3162A,2019MNRAS.483.5188C} -hereafter A19 and C19- for a description).
We employed {\itshape MUSE scibasic} and {\itshape MUSE scipost} routines to perform standard calibration steps to the raw data: master bias, (initial) master flat-fielding, twilight, illumination corrections; wavelength and flux calibrations. The datacubes for each exposure were reconstructed using the geometry and astrometry tables to a common 3D grid with a sampling of 0.2$\textrm{\arcsec} \times$\,0.2$\textrm{\arcsec} \times$\,1.25\,$\mathrm{\AA}$.
Subsequently, after refining the astrometry solution with a custom Python script, we improved the pipeline flat-field correction using the self-calibrating approach of the {\ttfamily CubeFix} routine, part of the {\ttfamily CubEx} package. 

Then, we performed the (flux-conserving) sky subtraction using the {\ttfamily CubEx} routine {\ttfamily CubeSharp}. These steps ({\ttfamily CubeFix} and {\ttfamily CubeSharp}) were repeated twice to minimize possible contamination by continuum sources. Finally, we employed 3$\sigma$ clipping algorithm to obtain the final average-combined J0952+0114 datacube with {\ttfamily CubeCombine} (see also B16, M18, A19 and C19 for further details).

The MUSE reconstructed RGB image of the J0952+0114 field is shown in Fig.\,\ref{fig:RGB_spectrum} where the RGB channels correspond to the pseudo broad V$-$, R$-$, and I$-$band images. The position of the QSO is indicated by the red cross. In Fig.\,\ref{fig:RGB_spectrum} we also present the MUSE J0952+0114 spectrum.\\

\begin{figure}[hb!]
\includegraphics[scale=0.45]{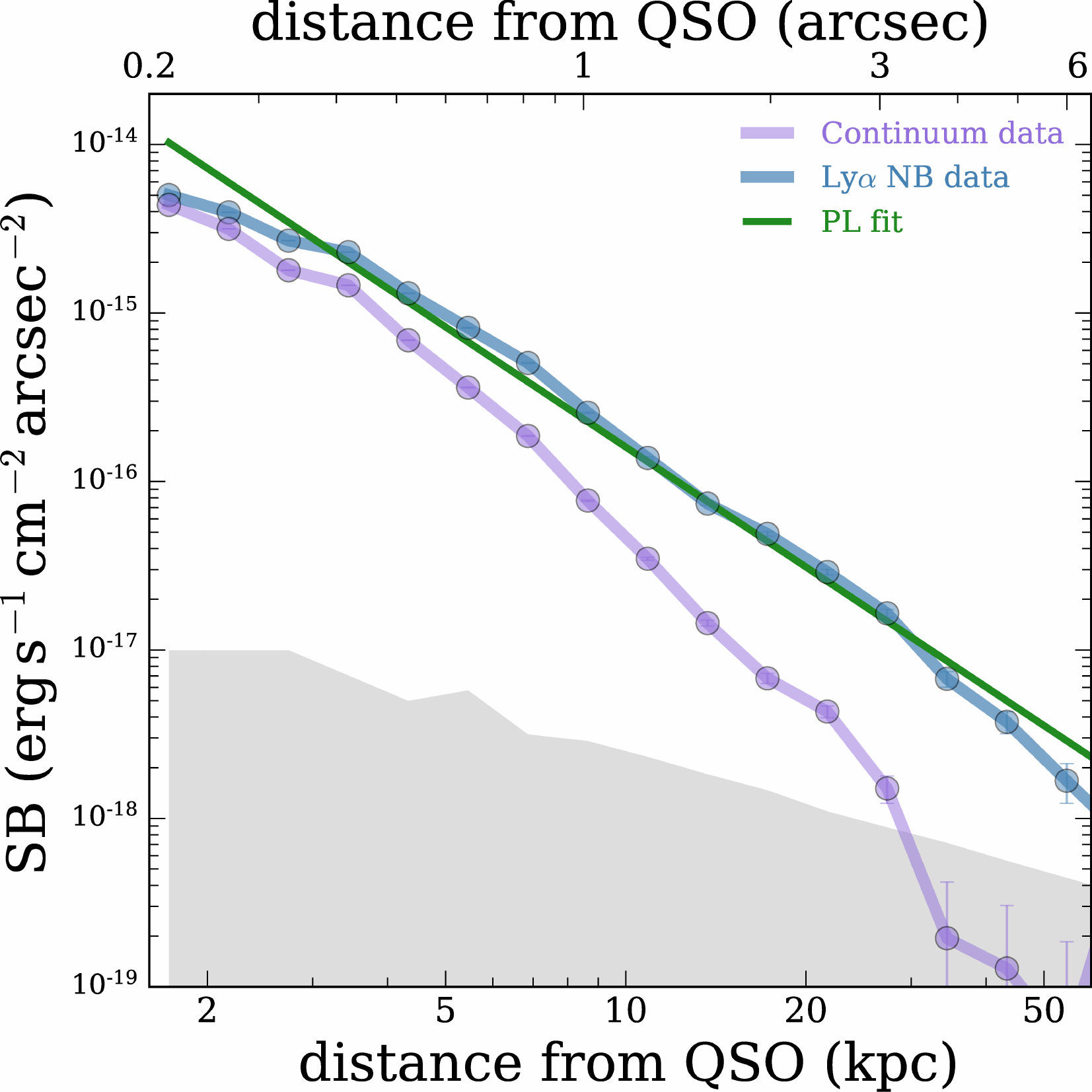}
\caption{ Ly$\alpha$ surface brightness (SB) profiles derived from the pseudo narrow-band (NB) Ly$\alpha$ image (only continuum subtracted) and continuum MUSE image. The circularly averaged profile of the continuum light is plotted in purple (solid line and circles) and the Ly$\alpha$ measurements in blue. The green line represents the power-law fit performed to the Ly$\alpha$ SB profile and the grey shaded area shows an estimate of the 2$\sigma$ gaussian noise associated with the Ly$\alpha$ SB profile.  \label{fig:SB_comp_Lya}}
\end{figure}

\begin{figure*}[ht!]
\centering
\includegraphics[width=0.32\textwidth]{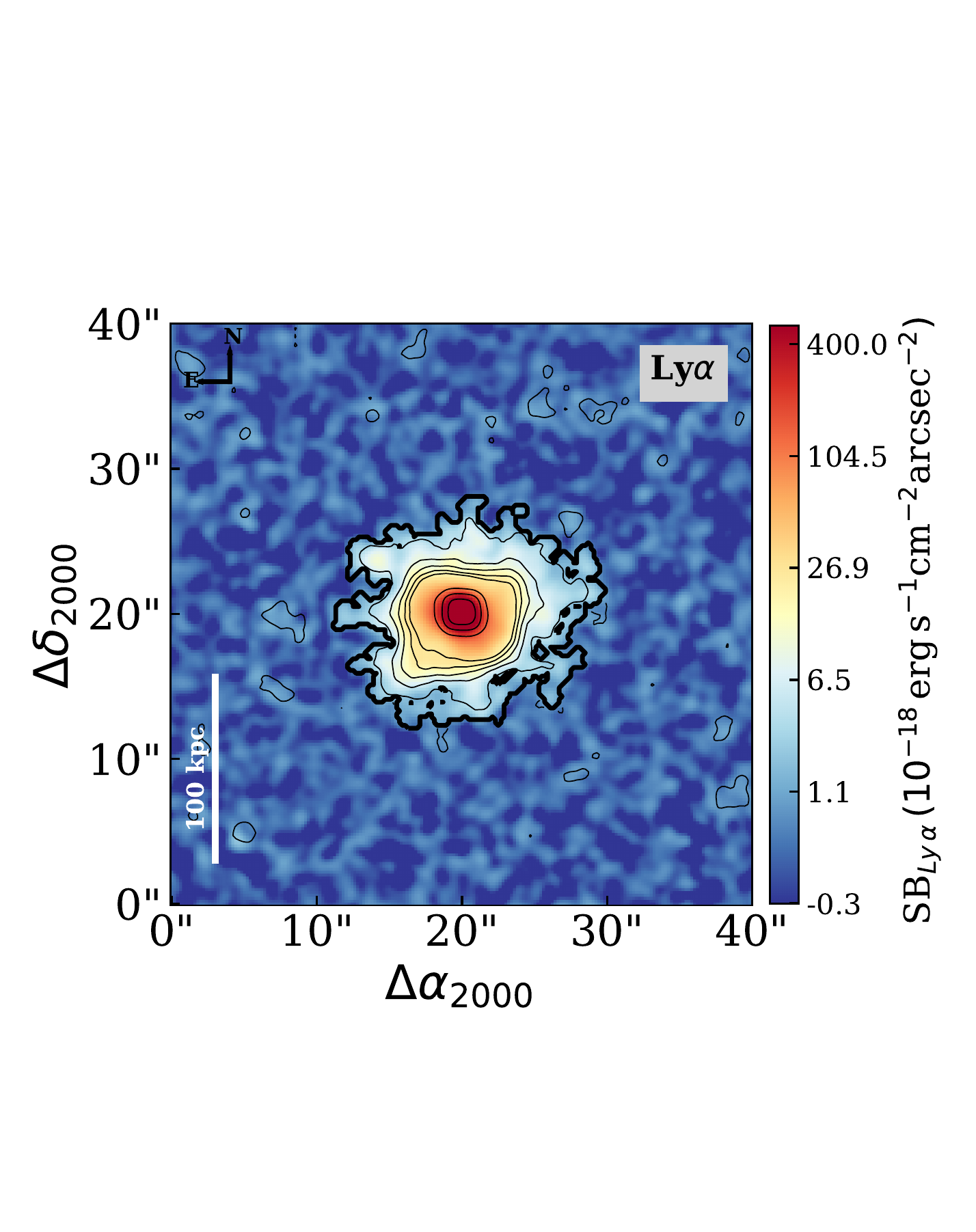}
\includegraphics[width=0.32\textwidth]{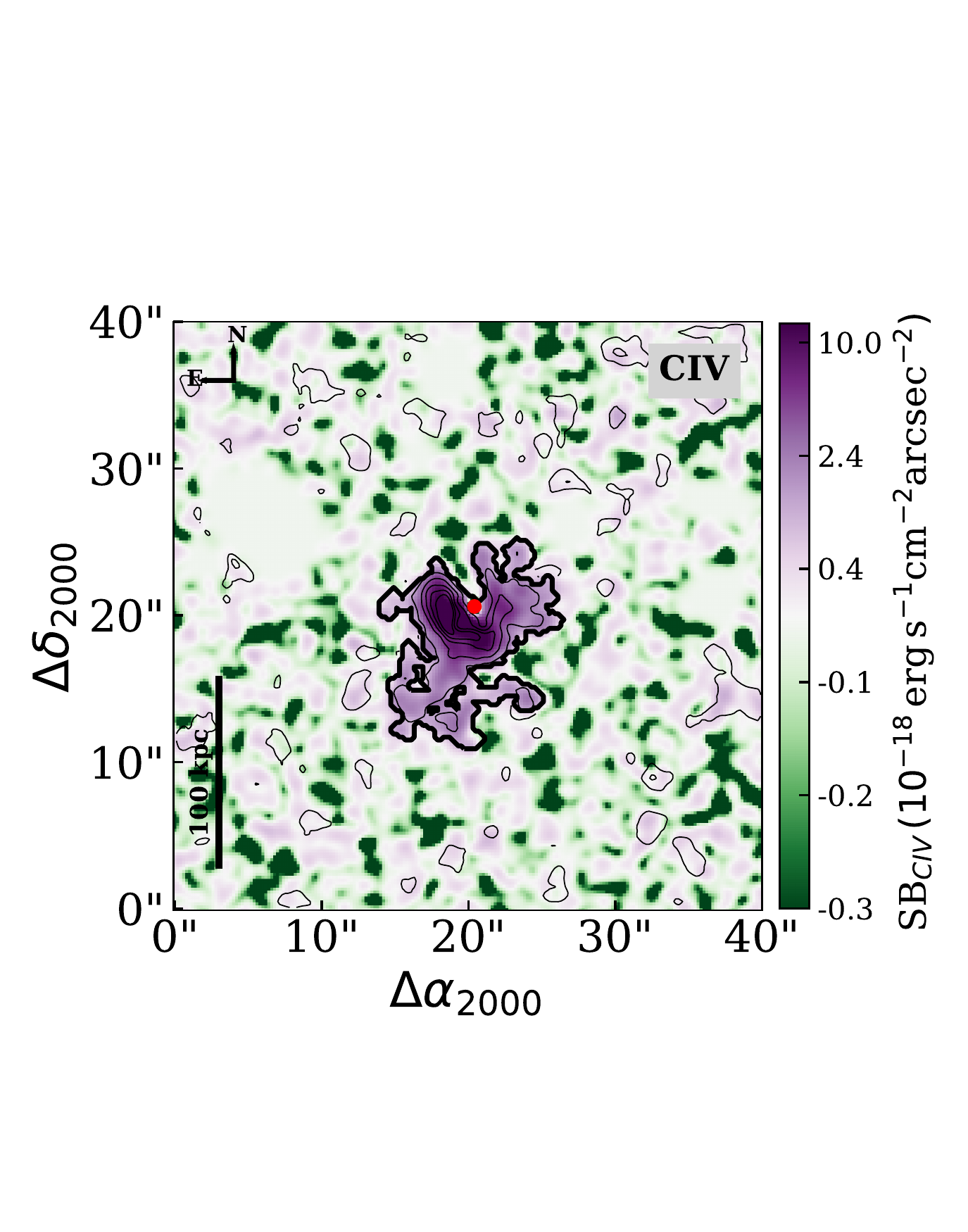}
\includegraphics[width=0.32\textwidth]{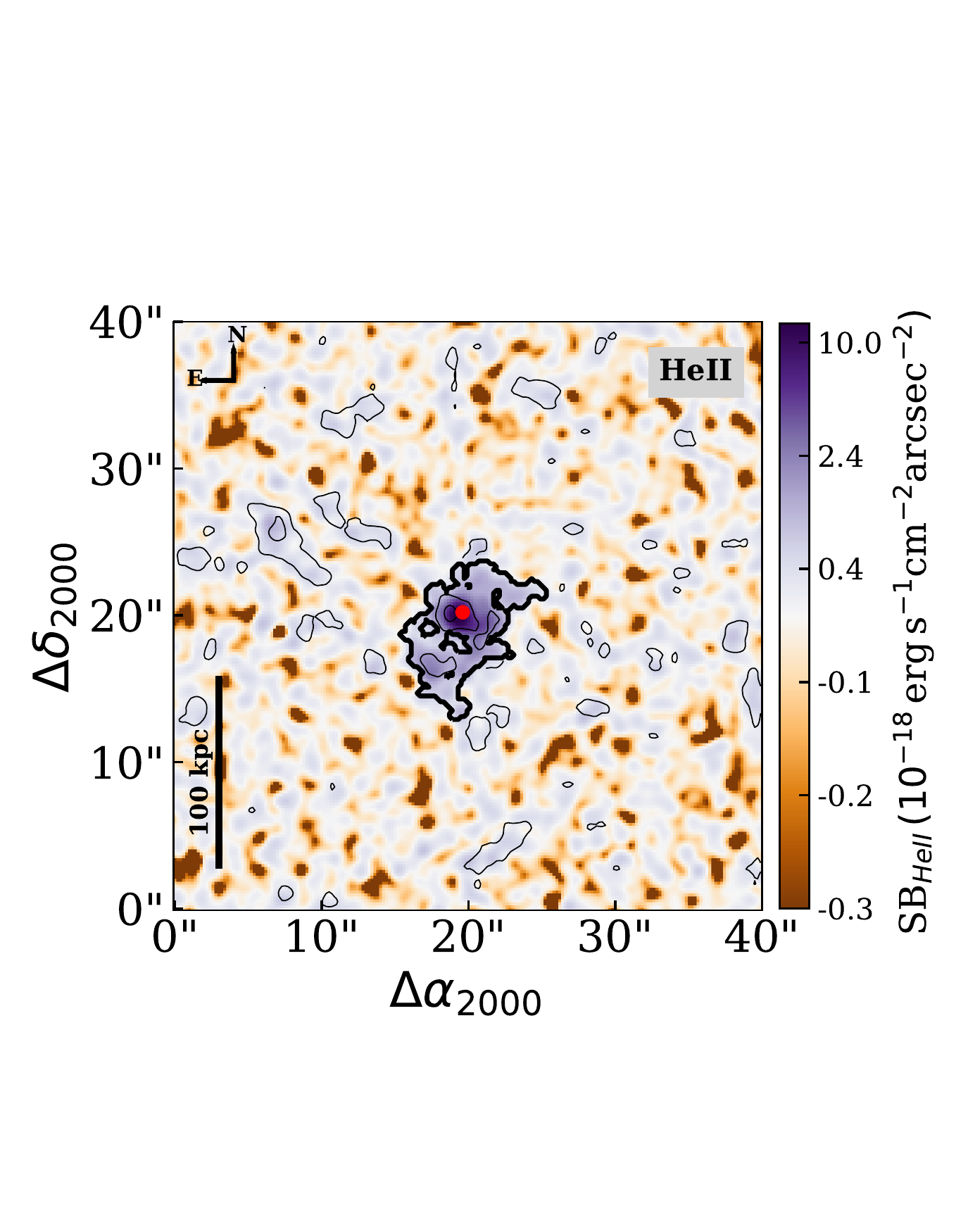}
\vspace{-1.2cm}
\caption{Ly$\alpha$\,$\lambda$1216, \ion{C}{4}\,$\lambda$1550 and \ion{He}{2}\,$\lambda$1640 emission line maps. These ``optimally extracted" maps are obtained from the PSF (only in the \ion{C}{4} and \ion{He}{2} cases) and continuum subtracted MUSE datacube with {\ttfamily CubEx}. Each map is the result of collapsing all voxels inside the 3D segmentation map along the wavelength direction. The black thick contour indicates the area detected above a S/N of 2 and corresponds to a SB of $\approx$10$^{-18}$ erg s$^{-1}$ cm$^{-2}$ arcsec$^{-2}$. The first thin contour represents a S/N of 4, while the following thin contours show increasing SNR levels with a step of 6. In each map, the area outside the thick contour represents a single noise layer added for visualization purposes. The red dot marks the QSO position. North is up and east to the left. Note the different sizes and morphologies of these emission line nebulae. \label{fig:nebulae}}
\end{figure*}

\section{Detection of the spatially extended emission} \label{sec:analysis}

The main aim of our analysis is to exploit the MUSE IFU's capabilities in characterizing the extended emission originating from the gas surrounding J0952+0114. 
The standard approach to isolate this ``pure" extended emission from that coming from the QSO is to 1) remove the QSO contribution (PSF subtraction) using the {\ttfamily CubePSFSub} routine part of the {\ttfamily CubEx} software, and 2) removal of contaminating background/foreground continuum sources using {\ttfamily CubeBKGSub}. This standard approach, which we used for the \ion{He}{2} and \ion{C}{4} emission lines, is described in more detail in B16, M18 and C19. We note that, in the case of the central Ly$\alpha$ emission for J0952+0114, the QSO Ly$\alpha$ emission is suppressed by the PDLA and, therefore, the QSO PSF subtraction may not be necessary for this line.\\

\begin{figure*}[ht!]
\centering
\includegraphics[scale=0.42]{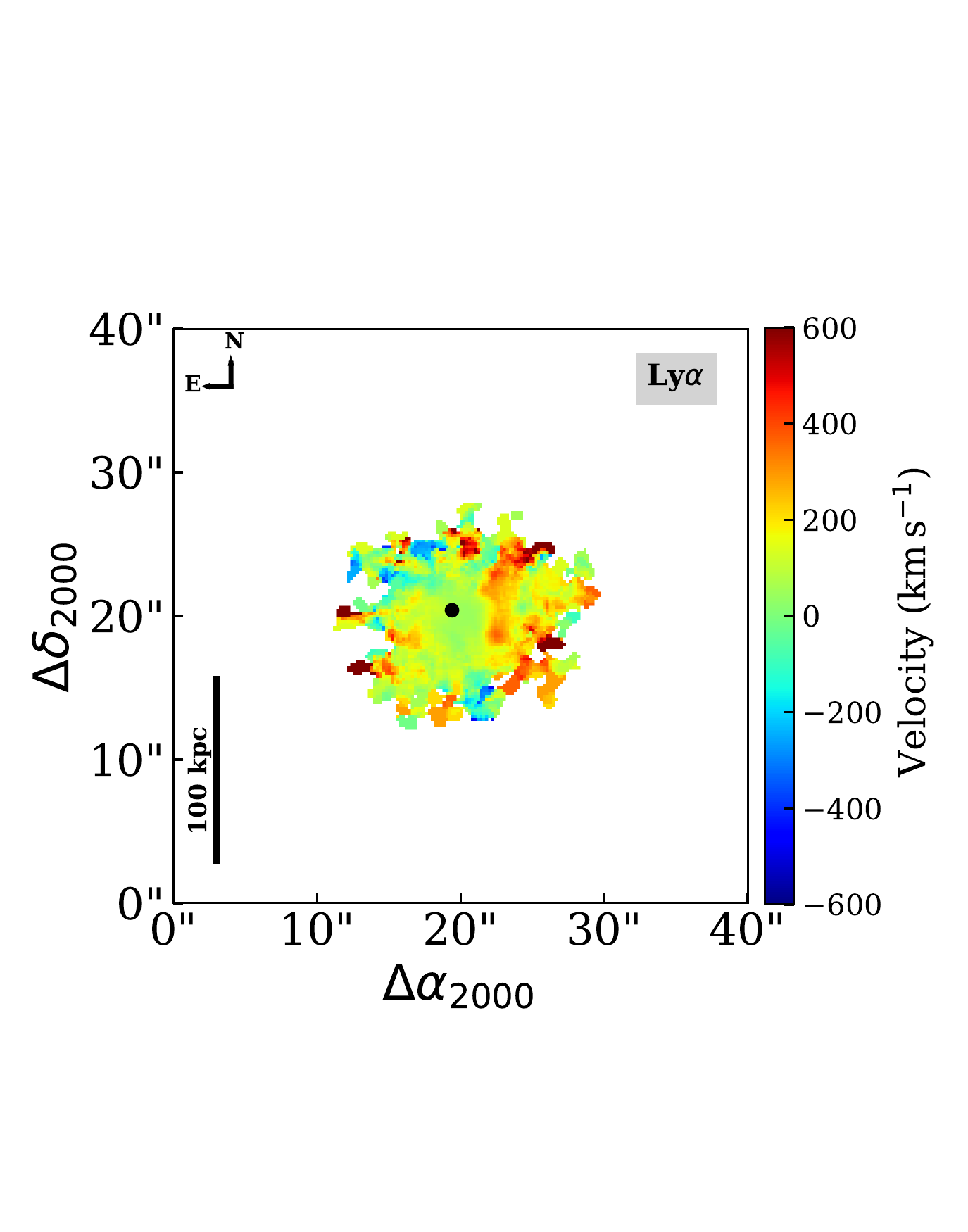}
\includegraphics[scale=0.42]{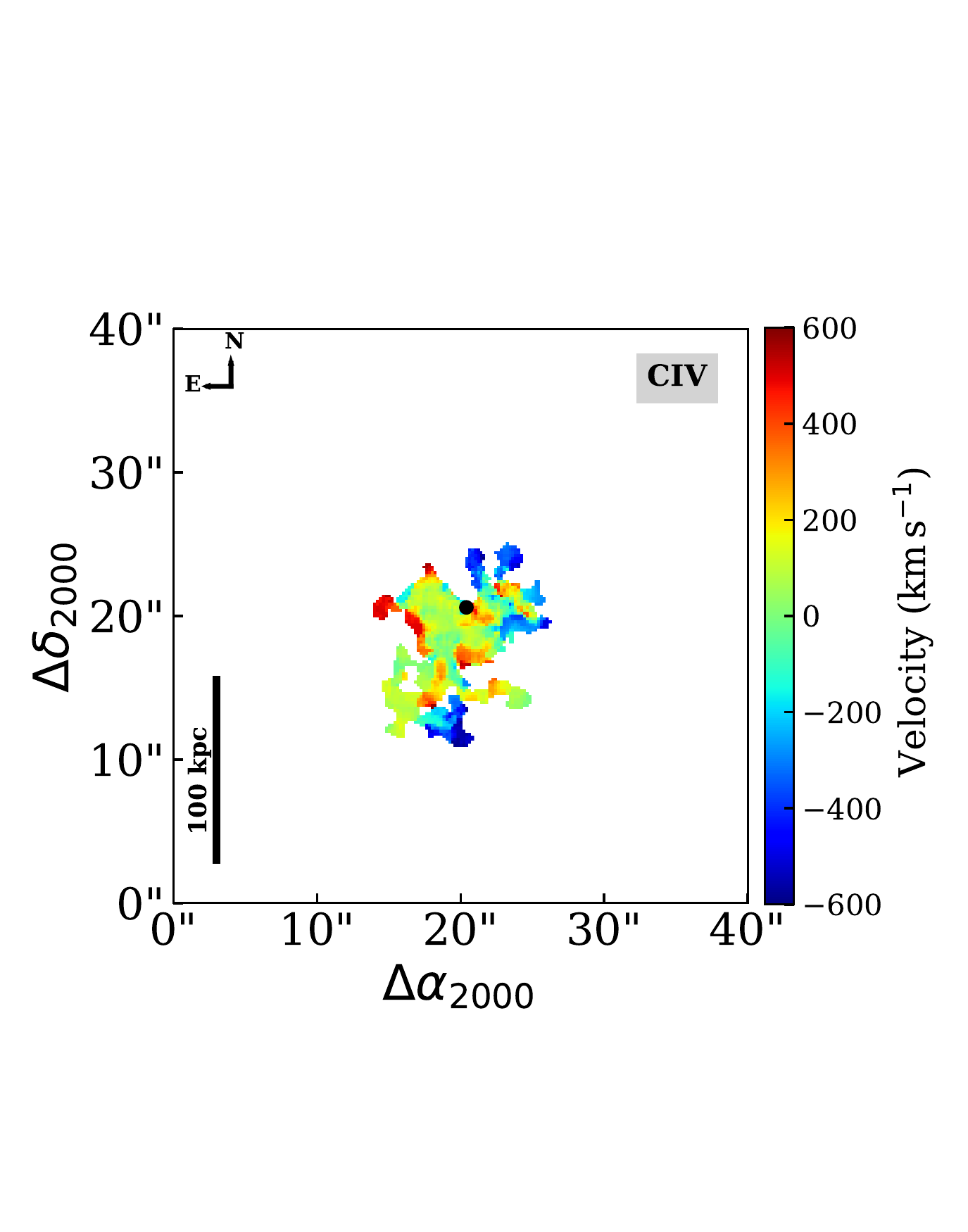}
\includegraphics[scale=0.42]{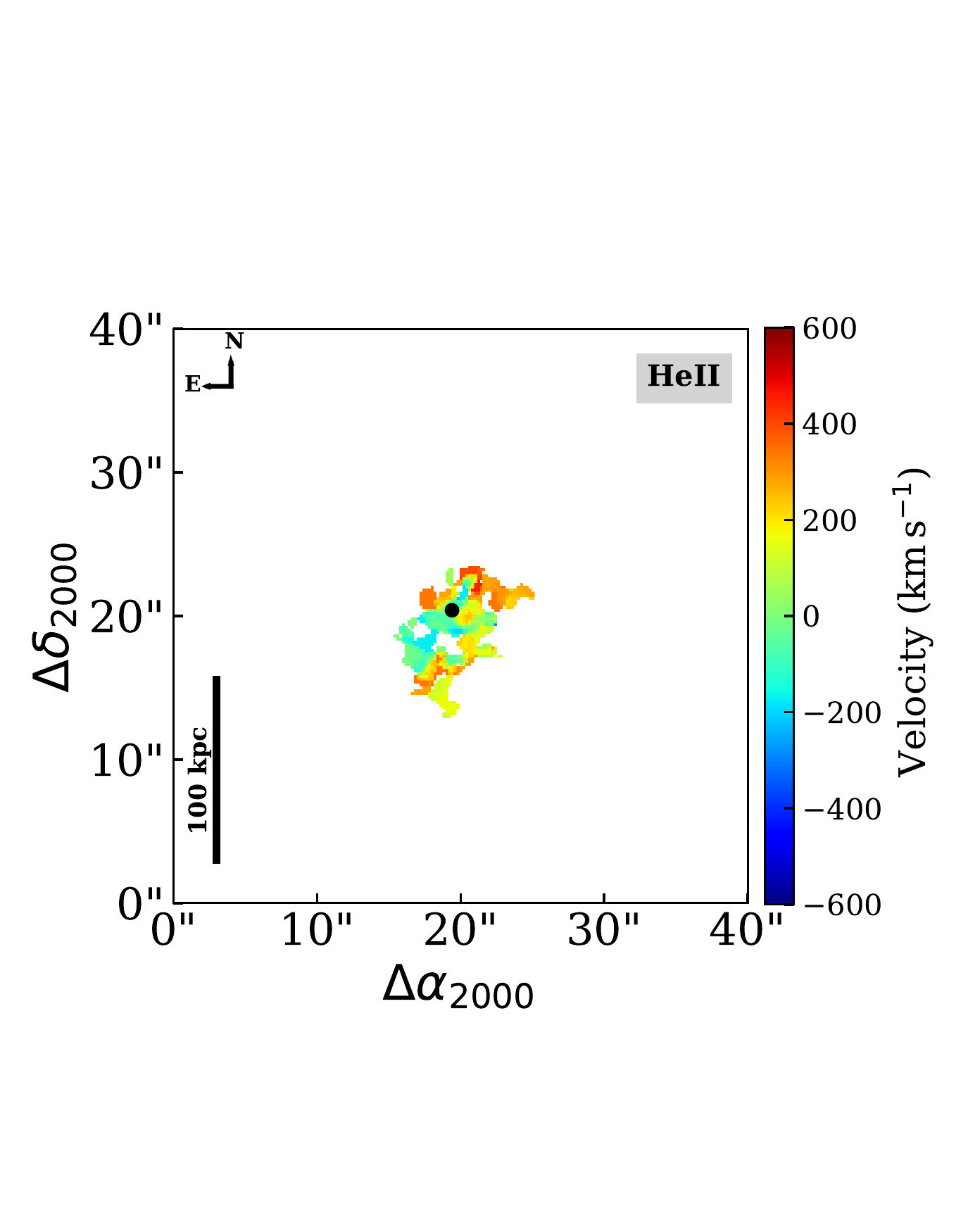}
\vspace{-1.5cm}
\caption{Bi-dimensional Ly$\alpha$, \ion{C}{4} and \ion{He}{2} velocity maps. These ``relative" velocity values are derived as the first moment of the flux distribution (in wavelength direction) within the 3D segmentation maps, giving an indication of the velocity centroid shift relative to the peak of the integrated flux. In each panel, the black dot marks the QSO position, the orientation of the map is north-east as in Fig.\,\ref{fig:nebulae}. On average, despite the wide dynamical range, all $\vec{v}$ maps do not show any clear kinematic structures. Only in the case of the non-resonant \ion{He}{2} $\vec{v}$ maps relatively higher velocities values are measured in the north-west tail.\label{fig:velocity}}
\end{figure*}

\subsection{Analysis of the extended Ly$\alpha$ emission}

We first tried to quantify the QSO contribution to the central Ly$\alpha$ emission considering the presence of both the peculiar narrow Ly$\alpha$ emission line shape and the PDLA. Due to the asymmetric profile and resonant nature of the Ly$\alpha$ line, performing any line fitting would not help in discerning between the contribution of the QSO and nebula to the integrated Ly$\alpha$ line flux. Therefore, we examined the Ly$\alpha$ surface brightness (SB) profile obtained from the continuum-subtracted but not PSF-subtracted Ly$\alpha$ narrow-band (NB) image. In Fig.\,\ref{fig:SB_comp_Lya}, the observed azimuthally averaged Ly$\alpha$ surface brightness profile is shown in blue and in purple is the continuum profile. We computed the Ly$\alpha$ SB profile from the pseudo-NB images by summing $\approx$ 30 $\mathrm{\AA}$ of the MUSE datacube in the wavelength dimension centered on the QSO Ly$\alpha$ emission. We also computed the continuum SB profile from the same MUSE datacube by integrating over 400 $\mathrm{\AA}$ redward of the \ion{N}{5}$\lambda$1243 line and rescaled by a factor of 40 to match the central NB SB profile.
Then, we performed both an exponential and power-law fits to the Ly$\alpha$ SB profile, considering up to a radius of 100 physical kpc (pkpc) and excluding the most internal part ($<$ 3 pkpc) close to the QSO position. We found that the Ly$\alpha$ profile is well traced by a power-law (with a slope $\alpha =$ -2.4, green solid line in Fig.\,\ref{fig:SB_comp_Lya}), suggesting that, if extrapolated to internal regions, the contribution of the extended nebula to the Ly$\alpha$ flux in the central region is substantial.
Therefore, the Ly$\alpha$ emission could be dominated by the nebula even in the region close to the QSO as also suggested by the relative excess of Ly$\alpha$ flux with respect to the continuum profile observed at the very smallest radii (beyond r\,$=$\,3\,kpc) and by the additional fact that the Ly$\alpha$ velocity map shows no feature at the center (see Fig.\,\ref{fig:velocity}). 
This is consistent with the PDLA absorbing completely or almost all of the intrinsic broad Ly$\alpha$ emission from the quasar.
We speculate that this may be the first case where the emission from a Giant Ly$\alpha$ nebula can be traced to such small radii, thanks to the serendipitous help of a natural \emph{coronagraph}.

Consequently, in the remaining analysis of the Ly$\alpha$ emission we neglect the contribution of the QSO (i$.$e$.$ a PSF subtraction is not necessary) and only consider the continuum subtraction. \\

\subsection{The \ion{C}{4}\,$\lambda$1550 and \ion{He}{2}\,$\lambda$1640 emission line detections}

In the case of the other emission lines (\ion{C}{4} and \ion{He}{2}), due to the clear broad component identified in the integrated spectrum (see Fig.\,\ref{fig:RGB_spectrum}), we followed the classical analysis procedure that includes both empirical PSF and continuum subtraction as already done in B16.
Briefly, our empirical PSF modeling is based on the estimation of the QSO flux in the central area of 1$\textrm{\arcsec}\,\times$ 1$\textrm{\arcsec}$ (or 5\,$\times$\,5 pixels$^{2}$, assuming that the central flux budget is dominated by the QSO) on pseudo-NB images of 187\,$\mathrm{\AA}$ wide (150 pixels in the spectral dimension ). Then, each reconstructed PSF image is rescaled using an averaged-sigma-clip algorithm and subtracted from the corresponding wavelength layer. After masking the spectral layers associated to the nebulae plus any continuum objects around the QSO, we iteratively ran {\ttfamily CubePSFSub} producing accurate results on large scales around the QSO as already successfully applied in B16, \citet{2017A&A...604A..23N}, and \citet{2018MNRAS.476.2421G}, among others.\\

\subsection{Extracting the extended emission lines in 3D}

Subsequently, we subtracted the continuum using {\ttfamily CubeBKGSub} based on a fast median-filtering approach with the most prominent emission lines masked. We optimized this procedure for each emission line. In particular, we created three different post-processed versions of the datacube around the most prominent emission lines: Ly$\alpha$, \ion{C}{4} and \ion{He}{2}.

\begin{deluxetable*}{l|c|c|c|c|c|c|c|c}
\tablecaption{Derived properties of the spatially extended and spectrally narrow emission lines.  \label{tab:el_table}}
\tablewidth{0pt}
\tablehead{
\colhead{Line} & \colhead{\# connected voxels}  & \colhead{$\lambda_{\mathrm{detected}}$\tablenotemark{a}}  & \colhead{Area\tablenotemark{b}} & \colhead{$\Delta\lambda$\tablenotemark{c}}   &  \colhead{Flux\tablenotemark{d}} & \colhead{L\tablenotemark{e}} & \colhead{$<\sigma>$\tablenotemark{f}}\\
\colhead{}    & \colhead{}    & \colhead{(\AA)} & \colhead{(arcsec$^{2}$)} & \colhead{(\AA)} & \colhead{(10$^{-17}$\,erg\,s\,$^{-1}$\,cm$^{-2}$)} &   \colhead{(10$^{42}$\,erg\,s\,$^{-1}$)} &  \colhead{(km\,s\,$^{-1}$)}
}
\startdata
Ly$\alpha$ & 33009 & 4882.00 & 176  &   37 &  742.8  $\pm$ 0.4 & 605.8  &  274$_{-120}^{+366}$\\
\ion{C}{4} & 11768 &        6218.95 & 83 &  30 & 46.2 $\pm$ 0.1 & 37.7 &  167$_{-42}^{+298}$\\
\ion{He}{2} & 2386 &        6589.27 & 41 &   19 & 9.6 $\pm$ 0.2 & 7.8 &  149$_{-40}^{+239}$\\
\enddata
\tablenotemark{}{}\\
\tablenotemark{a}{Flux-weighted wavelength.}\\
\tablenotemark{b}{The Area represents the number of pixels inside the 3D mask.}\\
\tablenotemark{c}{Size of the  3D mask spectral width. }\\
\tablenotemark{d}{The line flux is computed as the integrated value within the 3D mask.}\\
\tablenotemark{e}{The luminosity is derived assuming a redshift of $z$\,=\,3.02. }\\
\tablenotemark{f}{The $\sigma$ values are the median value of the bi-dimensional distributions shown in Fig.\,\ref{fig:dispersion} and the associated errors represent the 16th and 84th percentiles. For consistency with literature works, the gaussian-equivalent FWHM can be derived by multiplying $\sigma\,\times$ 2.355.}\\
\end{deluxetable*}


\begin{figure*}
\centering
\includegraphics[scale=0.42]{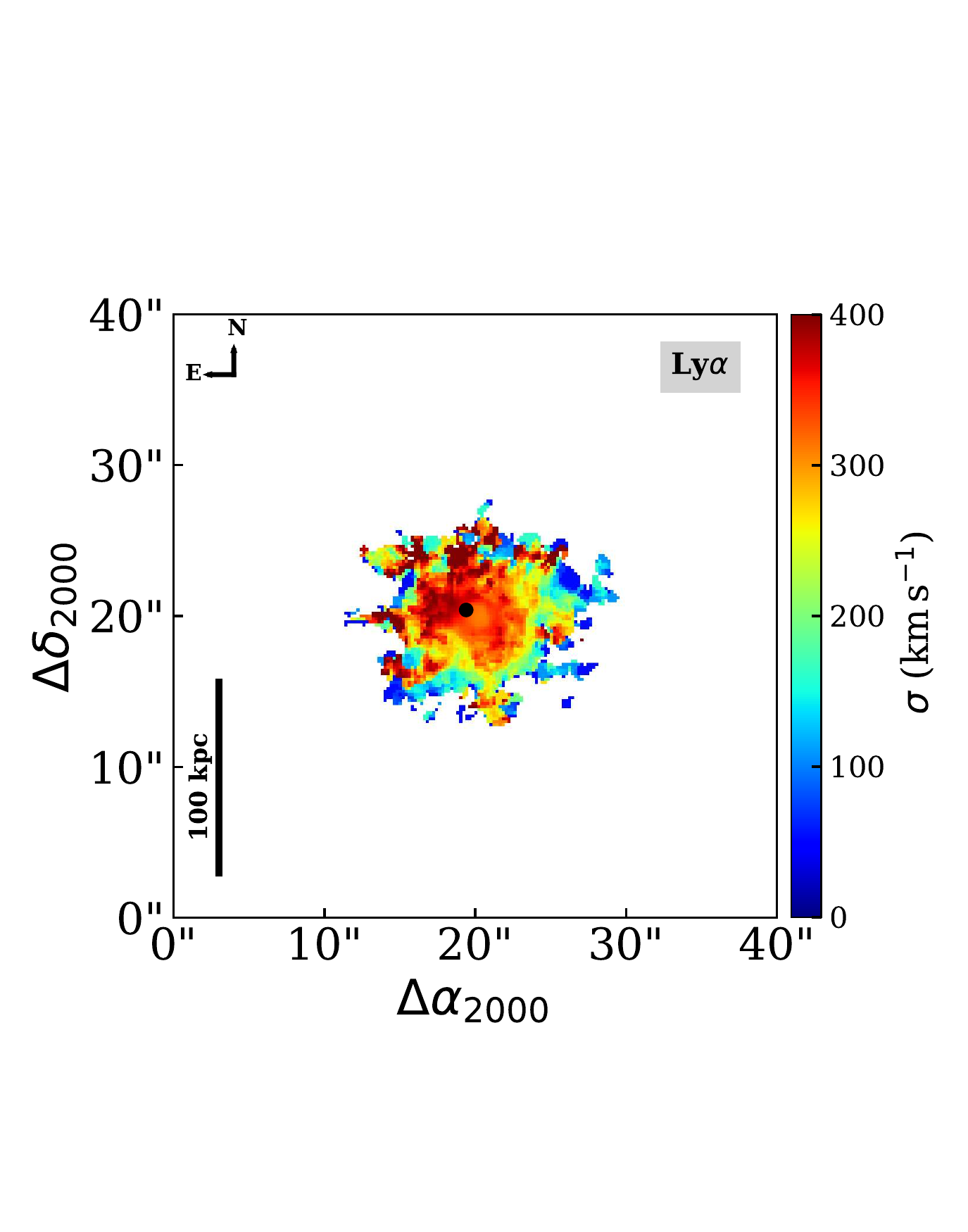}
\includegraphics[scale=0.42]{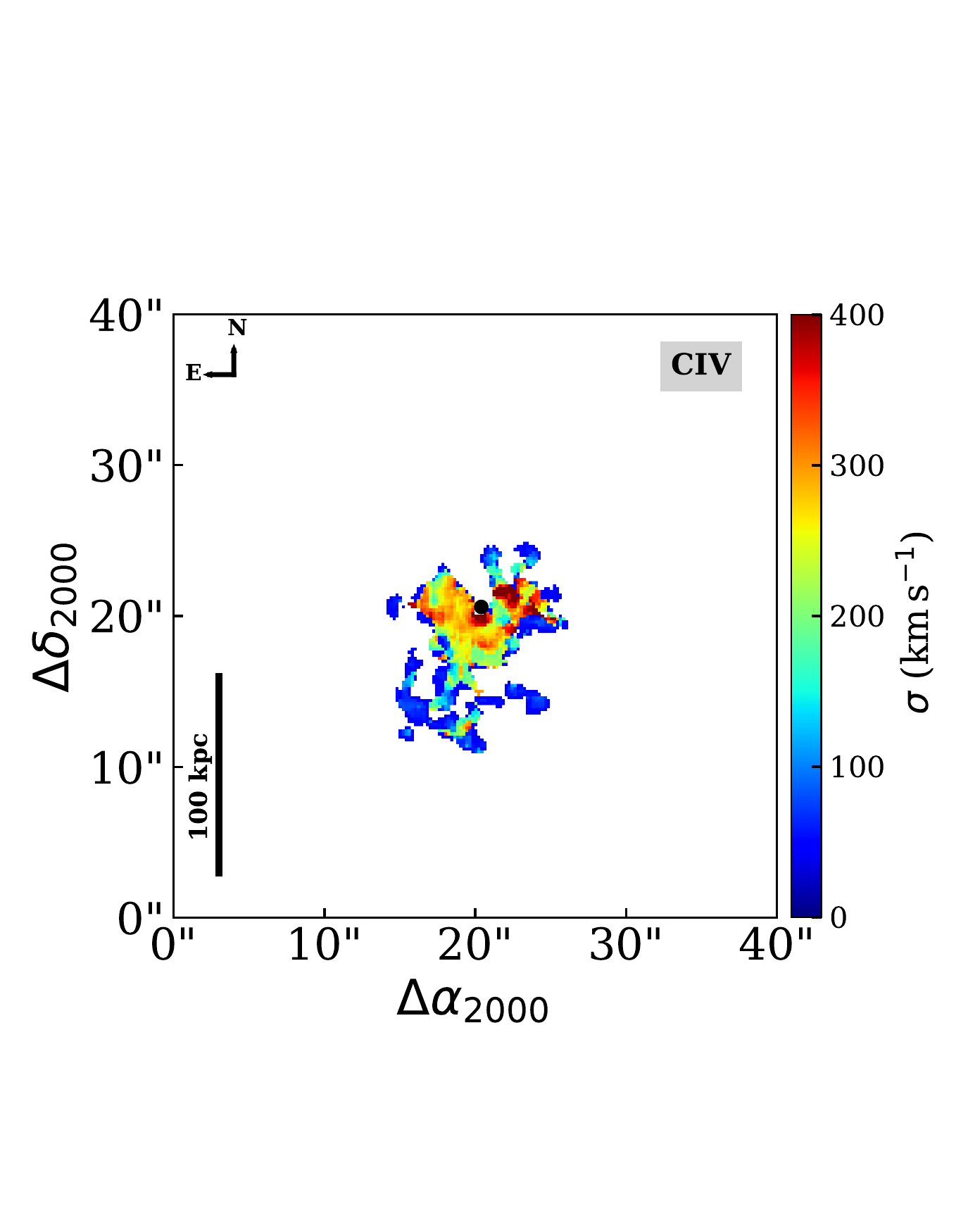}
\includegraphics[scale=0.42]{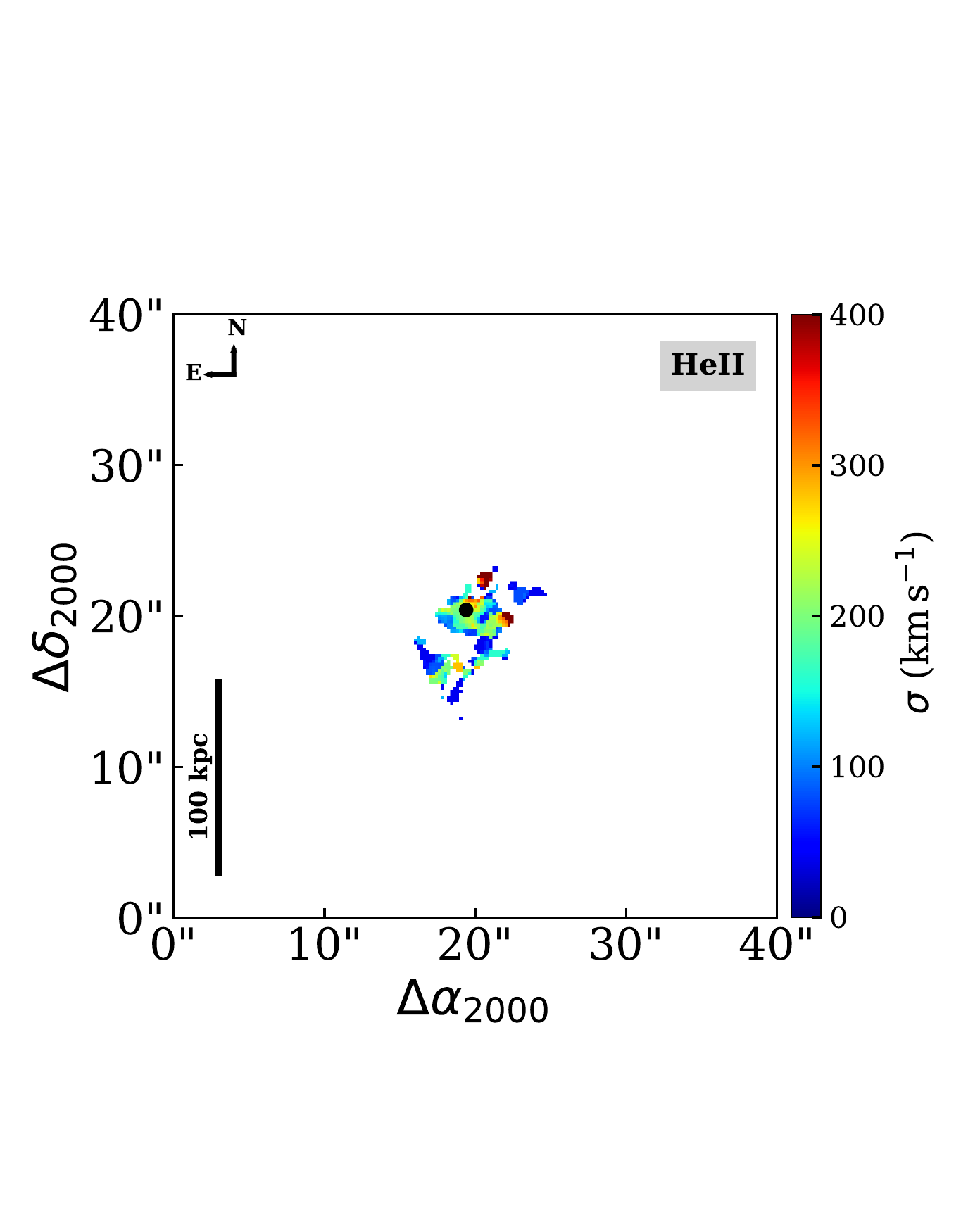}
\vspace{-1.5cm}
\caption{Bi-dimensional Ly$\alpha$, \ion{C}{4} and \ion{He}{2} velocity dispersion maps. The $\sigma$ values are obtained from the second moment of the same flux distribution used in the case of the $\vec{v}$ maps. There is a clear broadening of the Ly$\alpha$ line in the central region not detected in the other lines, possibly due to the radiative transfer effects. Both the \ion{C}{4} and \ion{He}{2} $\sigma$ maps present higher values in the inner regions with respect to the periphery of the nebula. In particular, we note the north-west ``arm" with high $\sigma$ values in the \ion{He}{2} map. As in the previous figure, the QSO position is indicated by the black dot and the maps are north-east oriented. \label{fig:dispersion}}
\end{figure*}


We selected post-processed sub-cubes, (80 $\mathrm{\AA}$ or 100 wavelength pixels wide) centered at the expected emission wavelength using as initial guess a QSO systemic redshift of 3.02 and searched for any possible extended emission. We then ran {\ttfamily CubEx} for object extraction and detection directly on the datacube above a user-defined signal-to-noise (S/N) threshold, in both the spatial and spectral dimensions.
In order to perform a consistent comparison between the different emission lines, we used {\ttfamily CubEx} with the most general input parameters as follows: i) (Gaussian) spatial filtering of 0.4$\textrm{\arcsec}$ (no smoothing in the wavelength direction); ii) spatial S/N ratio of 2 (equivalent to a SB of $\approx$10$^{-18}$ erg s$^{-1}$ cm$^{-2}$ arcsec$^{-2}$ in 1.25 $\mathrm{\AA}$) with a minimum of connected voxels\footnote{Spatial and spectral elements in the MUSE datacube.} of 1000 (after spatial smoothing); iii) rescaled variance; iv) masking of bright continuum sources and sky residuals.
In addition to Ly$\alpha$\,$\lambda$1216, we extracted both extended \ion{C}{4}\,$\lambda$1550 and \ion{He}{2}\,$\lambda$1640 emission lines. The 2D projection of the flux (Fig.\,\ref{fig:nebulae}), velocity (Fig.\,\ref{fig:velocity}) and velocity dispersion (Fig.\,\ref{fig:dispersion}) are based on the three-dimensional (3D-mask in short) segmentation mask obtained from this extraction procedure (see also B16 for further details). Table\,\ref{tab:el_table} summarizes the observed and derived physical properties for each emission line. The observed flux-weighted Ly$\alpha$ wavelength, 4882.0\,$\mathrm{\AA}$, corresponds to a redshift \textit{z}\,$=$\,3.0176.

\subsection{Analysis of the PDLA emission features}

\begin{figure}[]
\includegraphics[width=\linewidth]{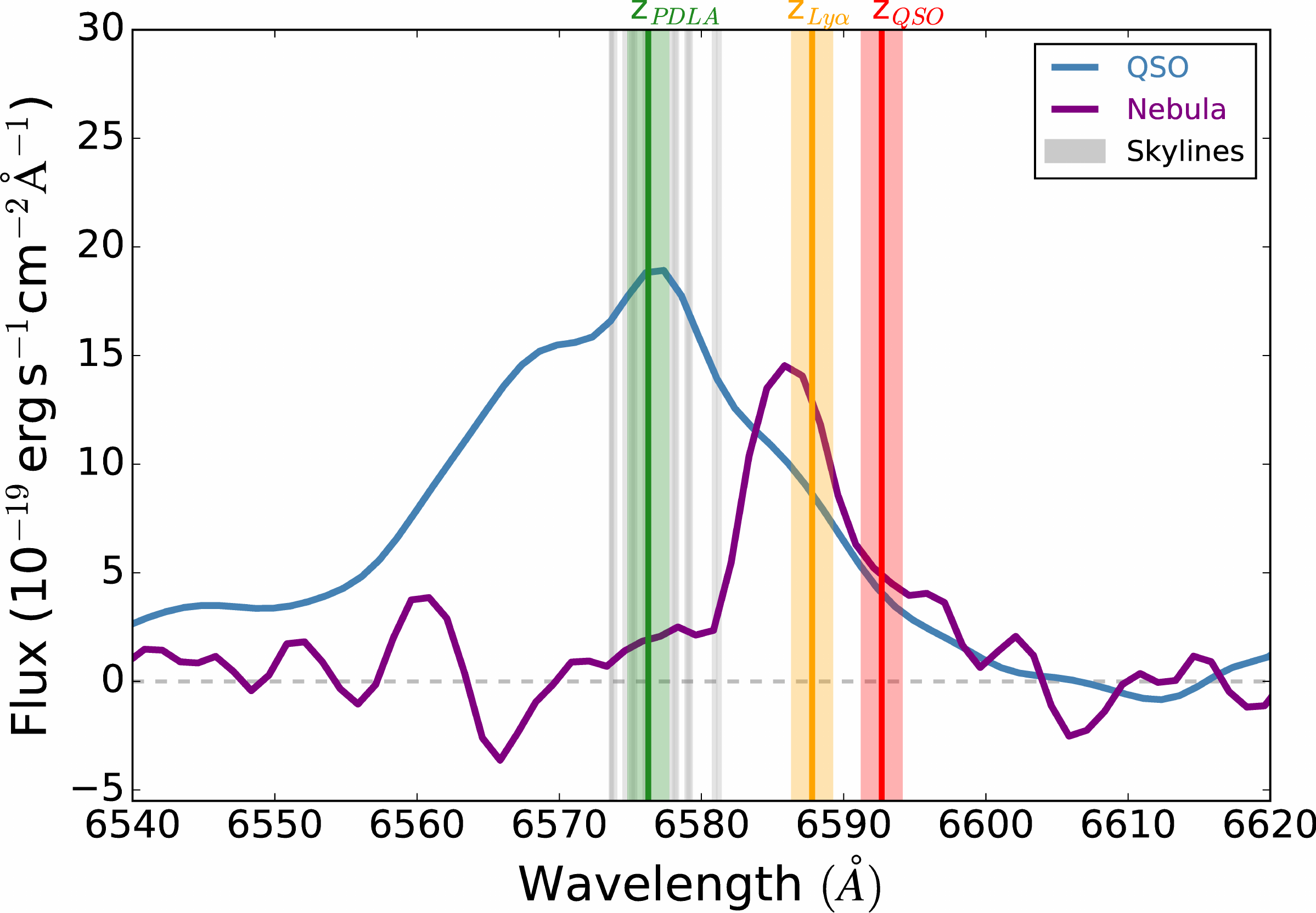}
\caption{Comparison between the \ion{He}{2} emission coming from the QSO (blue line) and nebula (purple line). The QSO integrated spectrum was obtained from a 0.6$\arcsec$ aperture centered on the QSO after the continuum subtraction was performed. The QSO spectrum was rescaled by a factor of 30 and a smoothing of 2 pixels (2.5$\mathrm{\AA}$) is applied for visualization purpose. In the case of the nebula, the purple spectrum represents the PSF subtracted \ion{He}{2} emission within a 1$\arcsec$ aperture shifted 2$\arcsec$ to the East direction from the QSO. The green, orange and red vertical lines indicate the \ion{He}{2} emission in case of adopting the PDLA, the Ly$\alpha$ or the QSO systemic redshift, respectively. The grey vertical lines are marking the position of the strongest residual skylines. \label{fig:HeII}}
\end{figure}

Previous studies (H04 and J16) confirmed the presence of a PDLA in the line-of-sight of the QSO J0952+0114 making use of the bluer wavelength coverage of the SDSS and BOSS spectra with respect to our MUSE dataset. From the analysis of the absorption lines (from Ly$\beta$ to Ly9) identified in the BOSS spectrum between 3600\,$\mathrm{\AA}$ and 4400\,$\mathrm{\AA}$, J16 inferred a redshift of $\textit{z}_{\mathrm{abs}}=$ 3.01 for the PDLA. Therefore, we analyzed our MUSE spectrum in order to find any possible emission and absorption features at the PDLA redshift. We caution that due to the small differences in redshift between the nebula (QSO) and the PDLA,  $\Delta \textit{z}=$ 0.007 ($\Delta \textit{z}=$ 0.01), and the higher intrinsic luminosity of the nebula (QSO) emission lines respect to the PDLA ones, possibly some of the components may be blended. We identified the PDLA emission lines in the continuum subtracted spectrum extracted from a region with a diameter of 0.6$\arcsec$ centered on the QSO. We found a clear detection of isolated \ion{He}{2} emission at the PDLA redshift (see Fig.\,\ref{fig:HeII}). We also note that for the \ion{C}{4} doublet lines there is a clear absorption system corresponding to the PDLA redshift. In the cases of the \ion{N}{5} and \ion{C}{3]} emission lines, we do observe some emission features but it is very difficult to distinguish between the contribution of PDLA and of the QSO/nebula to the global line profiles (see Fig.\,\ref{fig:PDLA_lines} in the Appendix).\\

\section{Properties of the spatially extended emission} \label{sec:results}
Taking advantage of MUSE's wide optical coverage, and after applying the methodology described above, we detected extended emission from the circumgalactic medium (CGM) of J0952+0114 in the Ly$\alpha$\,$\lambda$1216, \ion{C}{4}\,$\lambda$1550, and \ion{He}{2}\,$\lambda$1640 emission lines.
As described in Sect.\,\ref{sec:analysis}, all these lines are detected with a S/N ratio greater than 2 (after spatial smoothing per-voxel) and they have at least 1000 MUSE voxels connected (see Table\,\ref{tab:el_table}).\\

\subsection{Emission line Nebulae Morphology}\label{subsec:morph}

The ``optimally extracted" (OE) images of the emission line nebulae detected in the J0952+0114 QSO field at \textit{z}\,$=$\,3.02 are shown in Fig.\,\ref{fig:nebulae}. Differently from standard narrow-band (NB) images, we computed the OE maps using a differential spectral width for the nebulae and for the noise. 
By collapsing all the spectral layers detected in the 3D mask, features across a range of spectral widths can be highlighted with little noise. As a visual aid, we include a single layer of noise and the thick black contour in order to give a visual estimation of the noise level of our observed data.
Each OE map has a linear size of 40$\textrm{\arcsec}$, where the white (or black) vertical line shows a spatial extension of 100 pkpc and their orientation is North-East with the position of the QSO indicated by the red dot. The spectral width ($\Delta\lambda_{\mathrm{size}}$) of this OE map is specified in the fifth column of Table\,\ref{tab:el_table}.

It is worth noting that the detected emission lines present a large variety of shapes and sizes. For instance, the Ly$\alpha$ nebula extends for more than 130 pkpc\footnote{The nebula size is computed as the maximum projected sizes of the 3D masks} (1$\textrm{\arcsec}$ at \textit{z}\,$=$\,3.0176 corresponds to 7.6 pkpc) with a wide spectral width ($\Delta\lambda_{\mathrm{size}}$\,=\,30$\mathrm{\AA}$). In addition, the Ly$\alpha$ nebula has a more circularly symmetric appearance than the \ion{C}{4} and \ion{He}{2} nebulae (see Fig.\,\ref{fig:nebulae}). The integrated Ly$\alpha$ flux over the 3D mask is F(Ly$\alpha$) $\sim$ 7.4\,$\times$\,10$^{-15}$ erg s$^{-1}$ cm$^{-2}$, which, at this distance, corresponds to a luminosity of $\sim \, 6\, \times10^{44}$ erg s$^{-1}$, under the assumption of isotropic emission. 
Moreover, the \ion{C}{4} OE map (and marginally also the \ion{He}{2} OE map) presents a ``cavity" in the north direction that it is not observed in the case of the Ly$\alpha$ emission. The \ion{C}{4} emission map shows evidence of a clear asymmetry and a more filamentary structure with respect to the other emission lines. With a projected size of $\sim$ 100 pkpc and a total flux F(\ion{C}{4}) $\sim$ 5\,$\times$\,10$^{-16}$ erg s$^{-1}$ cm$^{-2}$ (i$.$e$.$ L(\ion{C}{4}) $\sim$ 4\,$\times$\,10$^{43}$ erg s$^{-1}$), it looks particularly elongated towards the south direction while in the north the aforementioned cavity is very prominent. The \ion{He}{2} extended emission maps, with a flux F(\ion{He}{2}) $\sim$ 9\,$\times$\,10$^{-17}$ erg s$^{-1}$ cm$^{-2}$, a luminosity of L(\ion{He}{2}) $\sim$ 8\,$\times$\,10$^{42}$ erg s$^{-1}$and a size of $\sim$ 75 pkpc, is characterized by a more compact morphology with a central circular component with two ``arms'' going out to the west and south directions.
Finally, we observe small spatial offsets between the peaks of the nebula emission and the QSO position, in particular, the offsets are larger than few arcsecs only in the case of the \ion{C}{4} OE maps. 
The difference in the emission line morphologies can be partly due to the different intrinsic surface brightness and therefore detection limits of different lines. It can, however, also tell us something about intrinsic gradients associated with different physical conditions of the ionized gas (i$.$e$.$ ionization parameter) within the nebula. 

The \ion{He}{2} (and \ion{C}{4}) emission is more elongated than that of Ly$\alpha$. As \ion{He}{2} is a non-resonant line and requires photons of energy $\gtrsim$54 eV (to produce doubly ionized He that then recombines to \ion{He}{2}), its morphology, better traces the regions with the high ionization levels. If this is true, the gas within the elongated \ion{He}{2} structure either has a lower density than the surroundings, or it is subject to a more intense ionizing radiation field. The latter option, in particular, could be an indication that the bright \ion{He}{2} emission traces, to some extent, the ionization cone of the QSO. A similar explanation may also hold for the morphology of \ion{C}{4}, which is also bright in the region orthogonal to the main emission `cone' of the QSO as possibly traced by high ionization non-resonant lines (i$.$ e$.$ \ion{He}{2}).\\


\begin{figure}[hb!]
\includegraphics[scale=0.45]{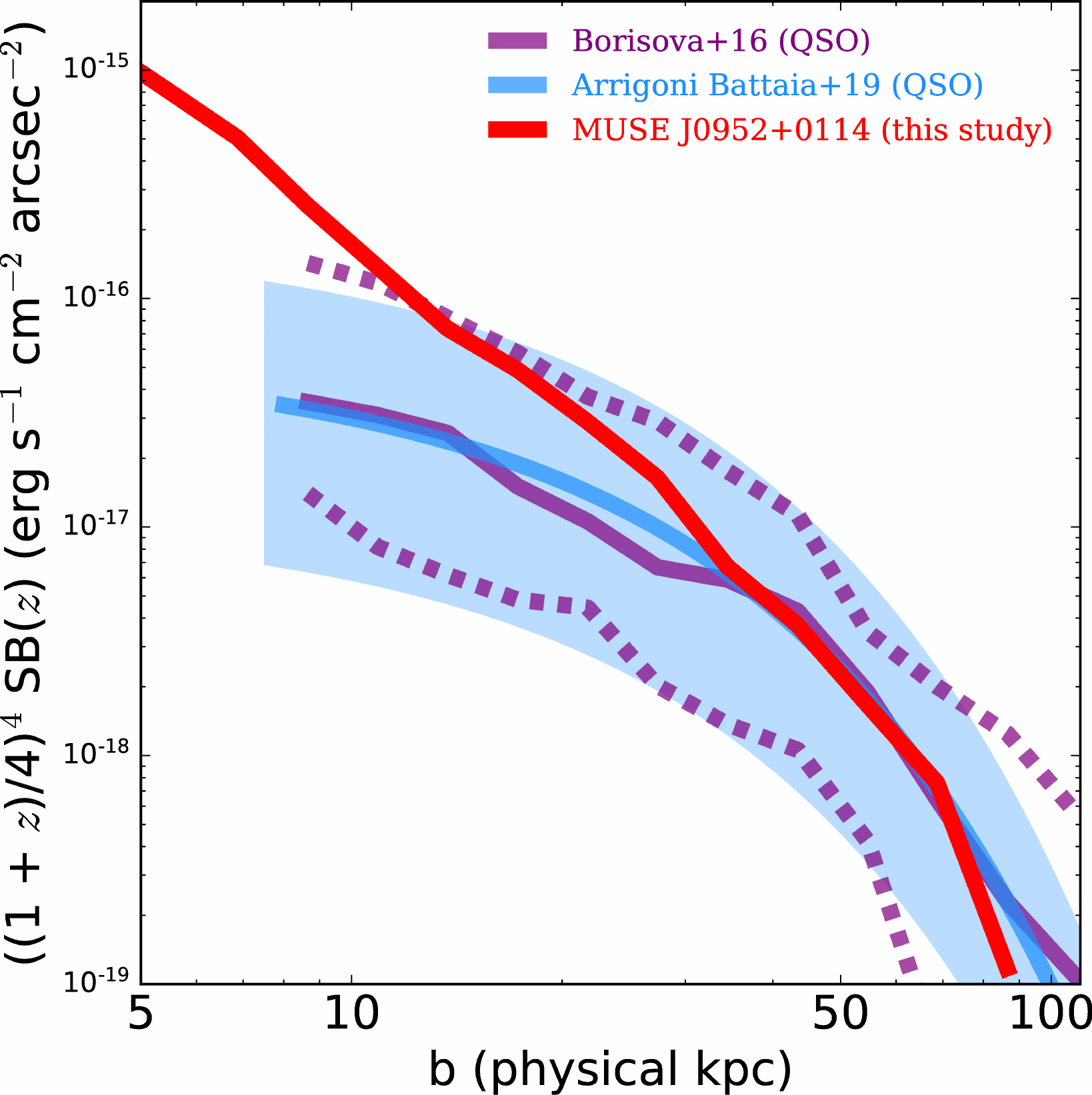}
\caption{ Ly$\alpha$ surface brightness (SB) profile of the J0952+0114 nebula as a function of the projected physical distance from the QSO. The circularly averaged profile is plotted with the red solid line. Other Ly$\alpha$ SB profiles are also plotted for comparison. In particular, the median MUSE Ly$\alpha$ nebulae SB profile from B16 is shown in purple with the dashed lines marking the 10th and the 90th percentiles. This median profile has been rescaled to \textit{z}\,$=$\,3. The best exponential fit (and its dispersion) of the SB profiles from the QSO MUSEUM is plotted with the solid blue line \citep[light blue area,][]{2019MNRAS.482.3162A}.\label{fig:SB}}
\end{figure}

\subsection{Ly$\alpha$, \ion{C}{4} and \ion{He}{2} Kinematics}

The resolved kinematic information is summarized in Figs.\,\ref{fig:velocity} and \ref{fig:dispersion} where we report the velocity, $\vec{v}$, and the velocity dispersion, $\sigma$, maps, respectively, in the case of the three extended emission lines detected in the J0952+0114 QSO field, i$.$e$.$ Ly$\alpha$, \ion{C}{4} and \ion{He}{2}.
The goal here is to identify any possible kinematical features, such as rotation, inflow, outflow or some special kinematic region within the nebulae. In particular, from the information given by the joint analysis of the 2D $\vec{v}$ and $\sigma$ maps for the three emission lines, the intriguing question hypothesized by J16 on whether there is evidence of an outflow at large radii from the QSO can be addressed.
It is well known that the Ly$\alpha$ line is not the best kinematic tracer given its resonant nature and some caution also has to be taken when deriving conclusion from the \ion{C}{4} line, which is also resonant.
As mentioned in Section\,\ref{sec:analysis}, we made use of the 3D segmentation masks to derive the first and second moments of the flux distribution (i$.$e$.$ we do not perform any fit to the emission lines) that give us an idea of both the velocity centroid with respect to the systemic redshift of the nebula (\textit{z}\,$\sim$\,3.017) computed from the flux-weighted Ly$\alpha$ wavelength (see Table\,\ref{tab:el_table}). We used this redshift as reference velocity for the other emission lines.

In general, we find that the dynamical range covered by the three lines in both $\vec{v}$ and $\sigma$ maps is in agreement with similar derived from other giant Ly$\alpha$ nebulae detected around typical bright QSOs without PDLAs. For instance, the $\sigma_{\rm{Ly}\alpha}$ is similar to the values measured by B16 and A19.
In the case of the Ly$\alpha$ nebula $\vec{v}$ map, we noticed some high velocity structure in the west direction. In the case of the \ion{C}{4} $\vec{v}$ map no clear kinematic patterns are found except for the lower velocity values in the periphery of the filamentary structure. Obviously, however, interpreting the Ly$\alpha$ line width in terms of intrinsic kinematics is not always an easy task, as the Ly$\alpha$ line can also be significantly broadened by radiative transfer effects. 

\begin{figure}[ht!]
\includegraphics[scale=0.8,width=\linewidth]{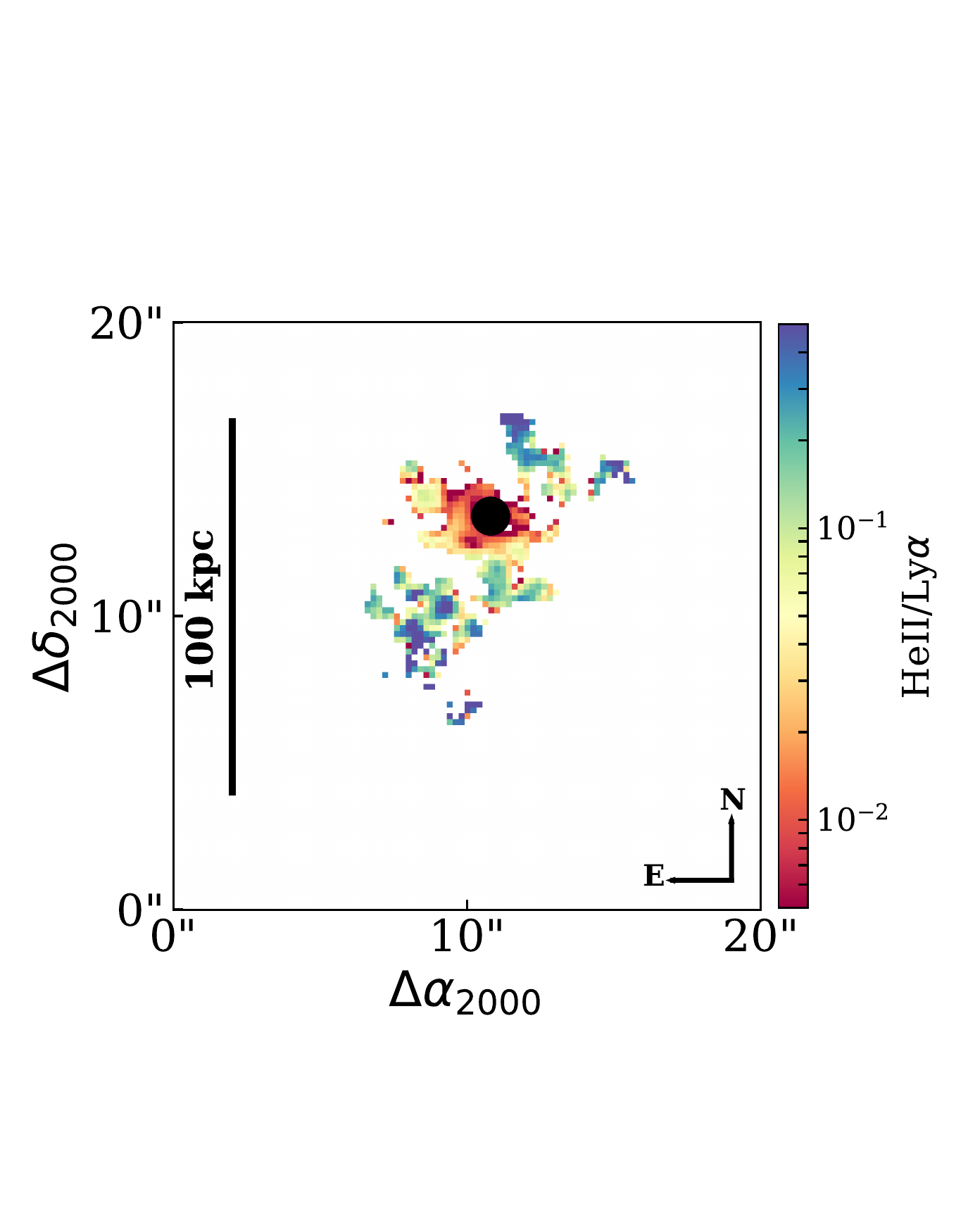}
\vspace{-2cm}
\caption{ Bi-dimensional \ion{He}{2}/Ly$\alpha$ ratio map. The ratio values are derived in the Ly$\alpha$, \ion{C}{4} and \ion{He}{2} common spatial mask (see text for details) from the smoothed (Gaussian smoothing with a radius of 2 spatial pixel) pseudo-NB images.  \label{fig:heii}}
\end{figure}

Moreover, the \ion{He}{2} $\vec{v}$ map shows a mildly coherent velocity gradient, with the arms (particularly the north-western) having higher values than the central region. Due to the poorer S/N of the arms with respect to the central region, only with deeper data would we be able to confirm this result. The velocity dispersion maps, shown in Fig.\,\ref{fig:dispersion}, look very different in the case of the Ly$\alpha$, \ion{C}{4} and \ion{He}{2}. A clear high velocity dispersion structure is detected in the Ly$\alpha$ and \ion{C}{4} $\sigma$ map in the central region. While the \ion{C}{4} and \ion{He}{2} $\sigma$ maps show a gradient with higher values in the center that decreases with the distance to the QSO.

Thanks to our MUSE data, on the one hand, we can confirm a fairly high $\sigma_{\rm{Ly}\alpha} \simeq 300 \; \textrm{km} \; \textrm{s}^{-1}$ as representative of a large portion of the nebula, even at spatial scales much larger than those observationally probed by J16. On the other hand, however, the velocity dispersion of the narrow non-resonant \ion{He}{2} line, which we can more safely take as indicative of the intrinsic dispersion of the cold CGM gas, is $\sigma_{\rm{He\,II}}$  $\sim 150 \; \textrm{km} \; \textrm{s}^{-1}$ (cfr.\ Figure\,\ref{fig:dispersion} and Table\,\ref{tab:el_table}). This is only one half of $\sigma_{\rm{Ly}\alpha}$ and smaller than the expected escape velocity. We conclude that the relatively large value of $\sigma_{\rm{Ly}\alpha}$ is likely not intrinsic and rather mostly a consequence of radiative transfer \citep{1991ApJ...370L..85N,2006ApJ...649...14D}.

\begin{figure}[ht!]
\includegraphics[scale=0.8,width=\linewidth]{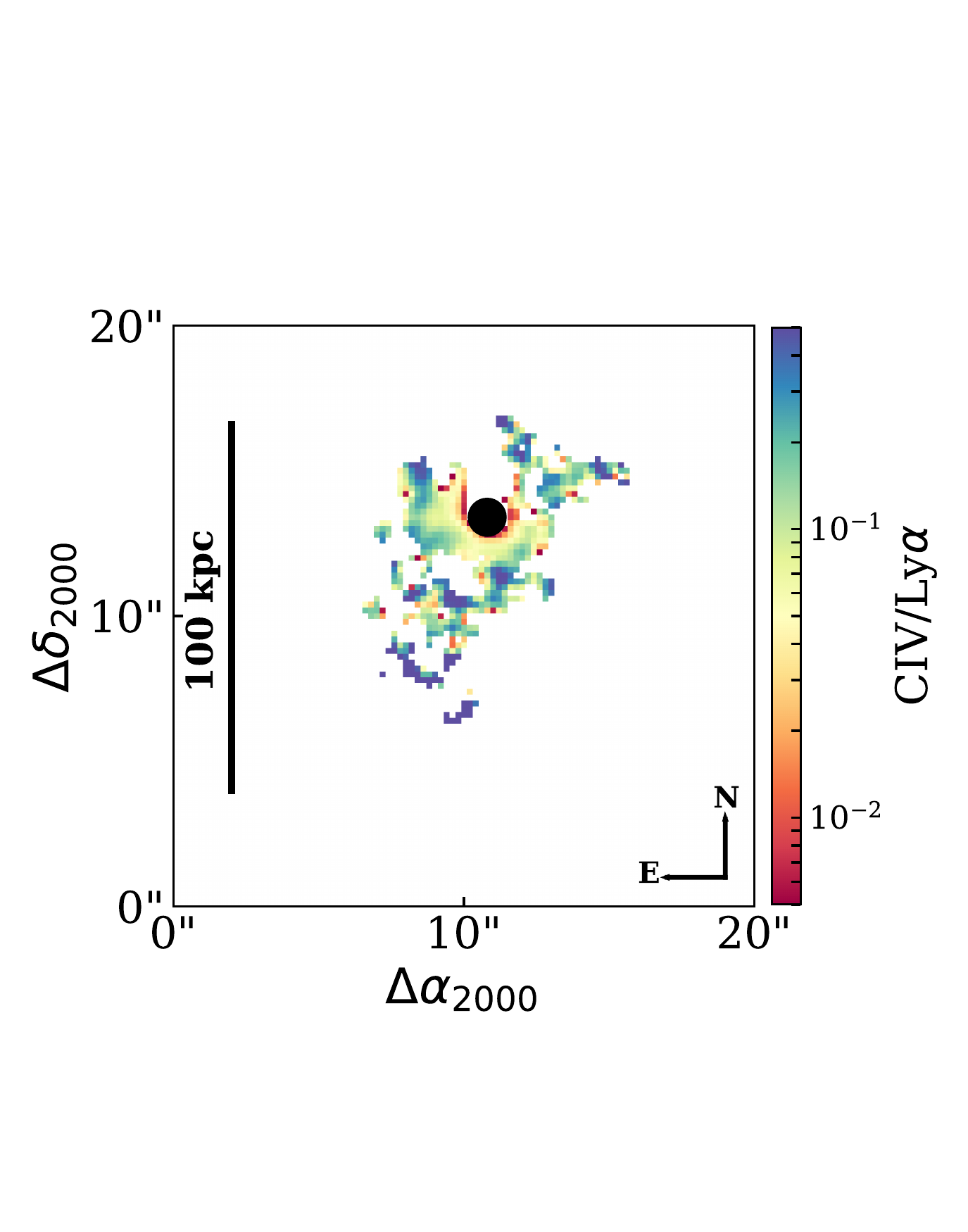}
\vspace{-2cm}
\caption{Bi-dimensional \ion{C}{4}/Ly$\alpha$ ratio map. As in Fig.\,\ref{fig:heii}, the line ratio is obtained from the smoothed \ion{C}{4} and Ly$\alpha$ pseudo-NB images. \label{fig:civ}}
\end{figure}

The velocity gradient in the \ion{He}{2} velocity map may also be interpreted as composed of two velocity structures, one in the center at lower velocities with respect to the two arm regions. Due to the significant heterogeneity and asymmetry of the velocity field, we rule out ordered rotation as a likely explanation for the observed kinematic pattern.  


Therefore, we conclude that we are most likely witnessing a significant gas circulation within the CGM of the galaxy hosting J0952+0114. However, we do not find convincing evidence for a strong outflow that is able to escape from the CGM in the central region from the \ion{He}{2} $\vec{v}$ and $\sigma$ maps (note also that the values in the very central region should be interpreted with caution due to possible PSF subtraction effects).


\subsection{Ly$\alpha$ Surface Brightness profile}
In Fig.\,\ref{fig:SB_comp_Lya} we show the Ly$\alpha$ SB profile of the nebula enshrouding the QSO J0952+0114. The PDLA blocks the QSO emission, and the extended nebula dominates the total Ly$\alpha$ emission even at small radii. Moreover, one of our objectives is to understand how the J0952+0114 Ly$\alpha$ nebula compares to other radio-quiet QSOs observed with MUSE.
In order to contextualize this nebula, we measured the SB Ly$\alpha$ profile through the standard NB images, similar to previous works (B16, A19). We computed the circularly averaged profile from the pseudo-NB image obtained collapsing the MUSE datacube using the maximum spectral size of 3D mask as the width, which, in the case of our Ly$\alpha$ emission is 30 $\mathrm{\AA}$. 
We then performed an exponential fit and a power\,-\,law fit above the 2$\sigma$ noise limit, i$.$e$.$ SB of $\approx$ 0.5$\times$10$^{-18}$ erg s$^{-1}$ cm$^{-2}$ $\mathrm{arcsec}^{-2}$ that returns a normalization parameter (at a radius of 10 pkpc) of log$_{10}$(C$_{p}$,r$_{10}$) $=\,-$15.8 and a slope $\alpha$ of $-$2.4, compatible with other Ly$\alpha$ SB profiles obtained with MUSE. More details on the fitting process are provided in Appendix B of B16.
The red line in Fig.\,\ref{fig:SB} shows the circularly averaged values computed in units of SB\,$\times$\,((1 + \textit{z})/4)$^{4}$ with an extension of more than 100 pkpc from the QSO. We chose these units because we wanted to perform a consistent comparison of the J0952+0114 SB profile with the other Ly$\alpha$ systems in the literature and, therefore, avoid any redshift dependence and cosmological dimming effects. Fig.\,\ref{fig:SB} also shows the comparison with previous works as the median profile of the MUSE QSO nebulae we published in \citealt{2016ApJ...831...39B} (purple line, see also Appendix C) and the QSO MUSEUM sample best exponential fit (A19, blue line).
As discussed later in Sect.\,\ref{sec:discussion}, the outer parts (beyond 10 pkpc) of the J0952+0114 SB profile are very similar to what we found in other giant fluorescent nebulae around QSOs at $z$\,$\approx$\,3. The innermost regions are brighter due to the presence of the PDLA, however, this could reflect the difficulty of tracing the nebula into the center of typical objects (without PDLAs).

\subsection{\ion{He}{2}/Ly$\alpha$ and \ion{C}{4}/Ly$\alpha$ map ratios}

Ly$\alpha$\,$\lambda$1216, \ion{C}{4}\,$\lambda$1550 and \ion{He}{2}\,$\lambda$1640 are some of the brightest rest-frame UV lines employed to better constrain the nature of ionized gas in emission.  In particular, the ratios of these lines, i$.$ e$.$ \ion{He}{2}/Ly$\alpha$ and \ion{C}{4}/Ly$\alpha$, are particularly useful for insights into the physical properties of the emitting gas, such as the ionization parameter, density and metallicity \citep[][C19 and references therein]{1997A&A...323...21V,2015ApJ...804...26A, 2015ApJ...799...62P}.

Differently from previous works, in the case of the gas surrounding J0952+0114, we were able to measure the spatial distribution 
of these line ratios instead of using limits. The resulting emission line ratio maps are presented in Figs.\,\ref{fig:heii} and \ref{fig:civ}. In order to avoid any aperture effects in the line ratios, we derived these maps with an intersection of the 3D segmentation masks (where we simultaneously detected the three emission lines), and smoothing of 2 spatial pixels (radius) for display purpose. 

The J0952+0114 field is a rare case among 
giant Ly$\alpha$ nebulae around high redshift QSOs in which these ratios can be measured in a consistent way and spatially resolved on extended scales. In general, the bi-dimensional distribution of \ion{He}{2}/Ly$\alpha$ and \ion{C}{4}/Ly$\alpha$ line ratios are not homogeneous, as also suggested by previous observations and theoretical works \citep[C19,][among others]{2018arXiv181105060C,2018MNRAS.473.5407M}.
Particularly, ratios in the inner region are lower and more ordered with respect to the outer parts where these values tend to be relatively higher and less spatially structured. In particular, using a 2$\arcsec$ region centered on the QSO, we measure a median \ion{He}{2}/Ly$\alpha$ ratio of 0.01 and a \ion{C}{4}/Ly$\alpha$ ratio of 0.05. If we move the same aperture in the outer part 1.6$\arcsec$ to the East and 4$\arcsec$ to the South of the QSO position, we obtain 0.09 and 0.11, respectively. In addition, the integrated values for both emission line ratios, measured over the entire maps, are \ion{He}{2}/Ly$\alpha$ $=$ 0.03 and \ion{C}{4}/Ly$\alpha$ $=$ 0.09, in agreement with previous MUSE studies of radio-quiet QSOs (see Fig. 8 in B16). Finally, the dynamical range covered by the resolved ratio maps confirms that the 2$\sigma$-limit values used in B16 were very conservative with respect to the detected values being a factor of two lower than the integrated measured ratios. This implies that the J0952+0114 nebula presents consistent emission line ratios compared to the MUSE radio-quiet nebula and is also compatible with some of the radio-galaxy haloes plotted in Fig$.$ 8 of B16.

The very low values of \ion{He}{2}/Ly$\alpha$ observed in the central region of J0952+0114 nebula, may be due to the fact that a significant fraction of the observed Ly$\alpha$ emission in these regions is not produced locally by recombination but due to scattering. We note however that, even in the most illuminated region, the observed \ion{He}{2}/Ly$\alpha$ ratio is mildly larger, but still about a factor of 2 below the expectation for pure recombination in a low density medium. This may imply that scattering effects are non-negligible in this region as well. 
At larger distances, the effect of the Ly$\alpha$ scattering should diminish, due to the fact that the optical depth to Ly$\alpha$ scattering can become sufficiently large and recombination radiation is expected to be the dominant emission mechanism (as in the case of the Slug nebula detailed in \citealt{2018MNRAS.480.2094L}). This would explain the higher line ratios measured in the outer regions.
Alternatively, the observed \ion{He}{2}/Ly$\alpha$ ratio could be an indication of the presence of unresolved high density clumps, or of a broad density distribution of the cold photo-ionized gas; this, in fact, has been proposed as a mechanism to explain low \ion{He}{2}/Ly$\alpha$ ratio in recombination-dominated regions of the CGM around QSOs. Indeed, at fixed average density, if the full density distribution is broader than a delta function, then some fraction of the gas can have a sufficiently large density that the helium is not completely ionized, with a consequent drop in the emissivity of the \ion{He}{2}\,$\lambda$1640 recombination line (see C19 for a more thorough discussion of this point).

With these caveats in mind, it is interesting to notice that the size of the observed nebula is up to one order of magnitude larger than those of the NLR models in \cite{2016MNRAS.456.3354F}, so, a first order (admittedly simplistic) scaling based on the ionization parameter would suggest densities down to 2 orders of magnitudes smaller than those of \cite{2016MNRAS.456.3354F}, $n_e \lesssim (1 - 10) \; \textrm{cm}^{-3}$, consistent with diffuse medium. Note that the simple scaling based on ionization parameter is a very crude approximation and that tailored photo-ionization models, considering the spatial dependence of the line ratios and radiative transfer effects, would be necessary to come to more secure conclusions.\\



\section{Discussion} \label{sec:discussion}

The new MUSE observations, described in Section\,\ref{sec:results}, add further, important elements to our understanding of the `exotic' quasar J0952+0114: 
\begin{itemize}
\item[] \textsc{i}) the quasar is surrounded by a giant Ly$\alpha$ nebula with SB profile and morphology similar to other nebulae discovered with MUSE (Fig.\,\ref{fig:nebulae});
\item[] \textsc{ii}) the central Ly$\alpha$ emission in the spectrum of J0952+0114 is consistent with the extrapolation of the SB profile of the giant nebula  (Fig.\,\ref{fig:SB}); 
\item[] \textsc{iii}) there are at least two components of \ion{He}{2} emission in the central part - a broader component at the redshift of the PDLA and a narrow component at the corresponding redshift of the Ly$\alpha$ emission (Fig.\,\ref{fig:HeII}).
\end{itemize}
In this section, we combine these elements with previous results and discuss the most likely scenarios for the origins of both the PDLA and extended emission around J0952+0114.  
Finally, we put these results into a broader context with a particular focus on AGN outflows.
 
\begin{figure}[]
\includegraphics[scale=0.5,width=\linewidth]{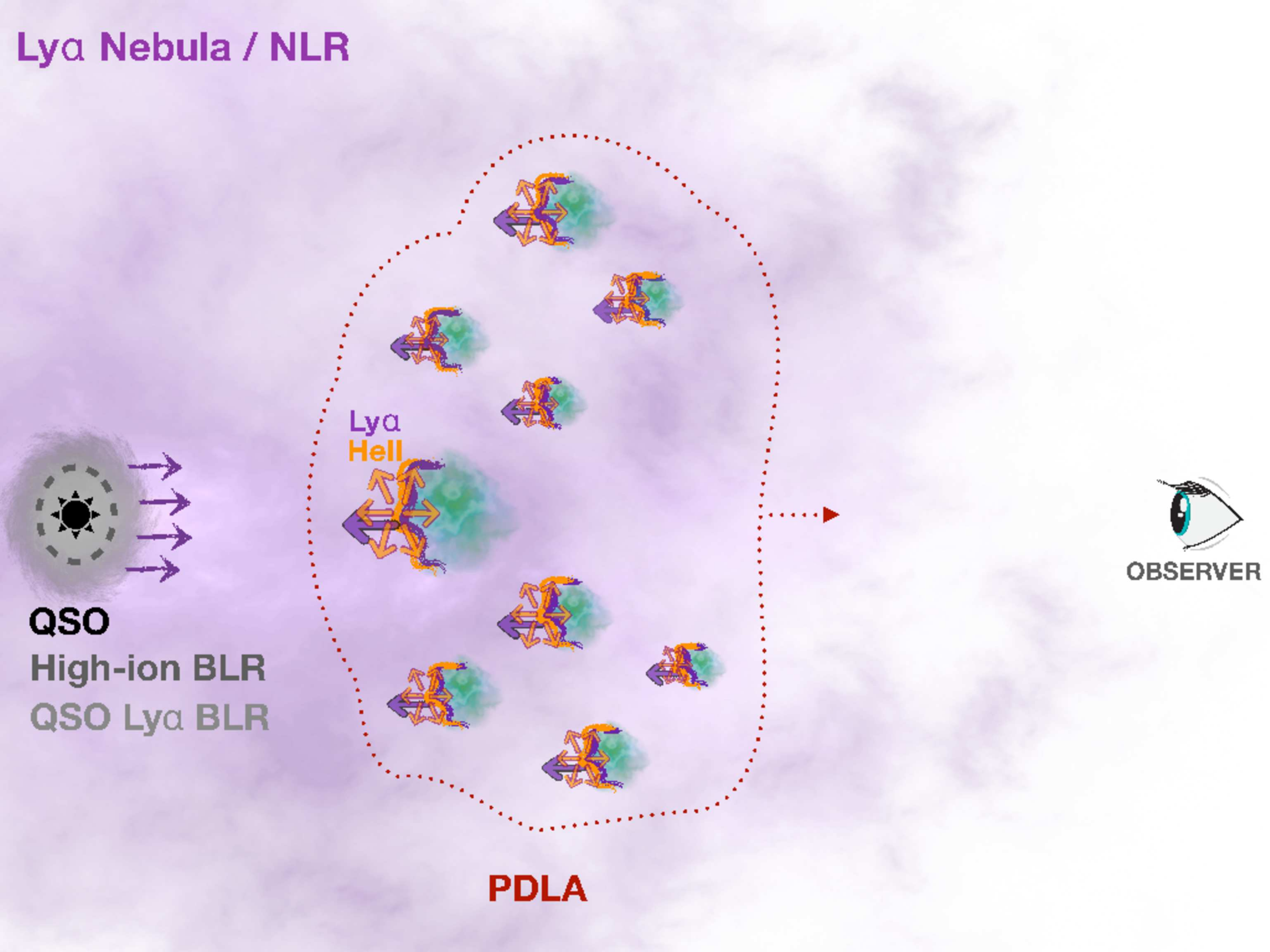}
\caption{Schematic view of the QSO J0952+0114 environment. The QSO at \textit{z} $=$ 3.02 is eclipsed by the PDLA composed of outflowing clouds (detected in absorption at \textit{z} $=$ 3.01) along the QSO line-of-sight. The Ly$\alpha$ emission from the BLR is emitted and scattered back toward the QSO (purple arrows) while the \ion{He}{2} recombination emission, due to its non-resonant nature, is able to pass through the PDLA clouds (orange arrows). The giant Ly$\alpha$ nebula and the NLR (in purple) are surrounding the QSO, the BLR and the PDLA.  \label{fig:cartoon}}
\end{figure}

\vspace{1cm}

\subsection{Origin and location of the PDLA}

The lack of observable Ly$\alpha$ emission from the broad line region (BLR), as already argued by previous studies, implies that the PDLA should have a size that is at least as large as the Ly$\alpha$ BLR of the quasar itself.
Recent reverberation mapping studies have found that BLR Ly$\alpha$ and  \ion{C}{4} emissions typically originate from similar regions \citep{2018ApJ...865...56L}. Since the  \ion{C}{4} from the BLR is visible in the spectrum of the quasar, this is suggesting that either the PDLA gas is not very optically thick to  \ion{C}{4} $\lambda$1550 (because of low metallicity or lower/higher ionization state) or that the part of the PDLA optically thick to  \ion{C}{4} is not fully covering the \ion{C}{4} BLR providing therefore a possible upper limit on the PDLA size.
The latter hypothesis would require the Ly$\alpha$ emitting BLR to be much smaller than the \ion{C}{4} one, which is in contradiction to the reverberation mapping results. Therefore, the analysis of the quasar spectrum alone does not provide therefore clear information on the size and spatial distribution of the gas associated with the PDLA.

Although indirectly, the presence and properties of the giant Ly$\alpha$ nebula discovered around J0952+0114 provide a constraint on the ``illumination" from the BLR on a much larger solid angle than our line-of-sight.
In particular, this applies to both the fluorescent recombination and BLR Ly$\alpha$ scattering case for the origin of the Ly$\alpha$ nebula (see \citealt{2017ASSL..430..195C} for a review).
Therefore, the presence of extended properties, i$.$ e$.$ the nebula, already excludes the scenario in which the PDLA fully covers the emitting solid angle of the BLR. 
On the contrary, the circular symmetry of the Ly$\alpha$ SB around the quasar, seems to indicate that the PDLA coverage is small and either uniform across the emission solid angle or restricted to a small cone that contains our line-of-sight below our spatial resolution scale. 
As shown in Section\,\ref{sec:results}, the giant Ly$\alpha$ nebula around J0952+0114 shares similar Ly$\alpha$ emission properties with respect to other nebulae detected around QSOs without PDLAs, reinforcing the idea that the majority of the emission solid angle is visible from the majority of line-of-sights through the CGM. 

This scenario has two possible implications: \textsc{i}) the PDLA is an intervening system in the CGM/ Inter Galactic Medium (IGM) (e$.$g$.$, gas associated with a foreground galaxy), not directly associated with the quasar, \textsc{ii}) the PDLA is intrinsically small and physically proximate to the quasar. Although neither of these two options can be discarded with certainty, we note that the broad \ion{He}{2} emission detected along our line-of-sight has exactly the same redshift as the PDLA (Fig.\,\ref{fig:HeII}). This  \ion{He}{2} emission is significantly different than what is typically observed in other quasars with similar luminosity, redshift, and UV slope (see Fig.s\,\ref{fig:MUSE_SDSS_BOSS} and \ref{fig:MUSE_SDSS_BOSS_zoom}) both in terms of equivalent width and shift from other high and low ionization lines, suggesting that it could be associated with the PDLA itself. If that is indeed the case, such broad and bright  \ion{He}{2} emission would require an ionization parameter high enough to imply that the PDLA should be possibly very close to the quasar, i$.$e$.$ within the Narrow Line Region (NLR), or certainly within the host galaxy. 
A narrow and fainter \ion{He}{2} component spectrally coincident with the extended nebula Ly$\alpha$ emission (and therefore redshifted by $\approx$ 500 km\,s$^{-1}$ with respect to the broad \ion{He}{2}) is also present close to the center suggesting the presence of a multi-phase medium with different kinematics and densities. In this scenario, the broad \ion{He}{2} component would represent therefore the non-resonant emission produced by a clumpy medium with a bulk velocity of $\approx$\,500 km\,s$^{-1}$ with respect to the host galaxy (which systemic redshift is taken from the narrow \ion{He}{2} component) and velocity dispersion of $\approx$\,600 km\,s$^{-1}$ (obtained from the FWHM of the broad  \ion{He}{2} component $\approx$ 1300 km\,s$^{-1}$). The narrowness of the metal absorption lines associated with the PDLA suggests that only one or a few of the clumps within this outflowing material is present along our line of sight. Because the broad \ion{He}{2} emission is spatially unresolved in our observations, this possible outflow should be confined on either nuclear or galactic scales. A schematic cartoon of the QSO J0952+0114 environment is presented in Fig.\,\ref{fig:cartoon}\footnote{This schematic view is not representative of the spatial physical scales, because we know that Ly$\alpha$ nebula has an extension of more than 100 pkpc and the size of the PDLA is not well constrained, therefore the goal of this cartoon is to give a closer view of the QSO and PDLA zone.}.

Such clumpy outflow would naturally explain all the features observed in both emission and absorption on both small and large scales but is of course not unique and, in particular, this scenario is degenerate with the possibility that the PDLA is intervening \emph{and} that the peculiar broad \ion{He}{2} emission is coming from the quasar BLR. In order to better disentangle this latter hypothesis, a statistical sample of quasar with similar spectra (see e$.$g$.$, \citealt{2015MNRAS.449.4204,2016ApJ...833..199J}) observed with MUSE could be used in future studies to verify the presence of other ``peculiar" broad \ion{He}{2} emission and confirm, at least in a statistical sense, the scenario suggested above. 

\subsection{The origin of the central Ly$\alpha$ emission}
Differently from the nebulae published in  \citealt{2016ApJ...831...39B}, in the case of J0952+0114, because the PDLA acts as a coronagraph, we have the unique capability of tracing the nebular emission to innermost QSO regions (as shown in Sec.\,\ref{sec:analysis}, we also caution about the fact that our data are seeing limited at the PSF FWHM $\approx$ 0.7$\arcsec$ level). Therefore, if the central, narrow and bright Ly$\alpha$ emission observed along our line-of-sight is not produced in the quasar BLR nor from the PDLA, what then is its origin? As we have argued in the previous section, the optically thick PDLA is only covering a small fraction of the quasar ionizing opening angle. As in the case of the illuminated CGM, the central component could indicate fluorescently emitting gas associated with: \textsc{a}) the Interstellar Medium (ISM) of the quasar host galaxy, \textsc{b}) centrally concentrated CGM material, \textsc{c}) some of the less dense material of the hypothesized outflow containing the PDLA. 
A close inspection of the velocity maps of both Ly$\alpha$ and narrow \ion{He}{2} emission in the central region as shown in Fig.\,\ref{fig:velocity} suggests that the latter possibility could be readily excluded because we do not see any change in velocities consistent with the PDLA blueshift in the central regions. However, we do see clearly detected extended  \ion{C}{4} emission, possibly suggesting that the inner regions have been enriched and possibly affected by (previous) outflow events.  Interestingly, such extended \ion{C}{4} emission suggests that the J0952+0114 nebula could be more metal-enriched than other MUSE QSO nebulae without PDLAs (B16, A19), and possibly providing a link between the outflow observed in absorption and the metal emission properties. In order to establish a possible correlation, it is necessary to build a larger sample of quasars with PDLA and Integral-Field observations as already discussed in \citet{2013A&A...558A.111F} and \citet{2018MNRAS.477.5625F}.
Regarding the possibilities \textsc{a} and \textsc{b}, we notice that both the SB profiles and velocity maps in the central regions do not show any abrupt changes (see Fig.s\,\ref{fig:nebulae} and \ref{fig:velocity}). These results suggest that the transition between CGM and ISM properties that determine the Ly$\alpha$ SB (e.g., density distribution and temperature) could be rather smooth or indistinguishable, at least in the region illuminated by the quasar. This is consistent with the recent results on the Slug Nebula (C19) that have suggested that the density distribution of illuminated ``cold" gas on CGM scales could be as broad as what typically observed in the ISM.  

\subsection{J0952+0114 and other AGN outflows}

QSO J0952+01144 is not the only case among observed high-redshift quasars for which possible signatures of outflows have been reported in the recent literature \citep[e.g.][among others]{2015MNRAS.446.2394B, 2017A&A...598A.122B, 2017A&A...601A.143F}.
Despite the large number of studies, however, little is know about the detailed properties of the various phases that could be associated with these AGN outflows. 
In this respect, it could be interesting to consider some analogies between the possible clumpy J0952+0114 outflow and the molecular outflows published in the recent works of \citealt{2015A&A...574A..14C, 2017A&A...608A..30F, 2018arXiv180600786B}. 
By using ALMA data (specifically the CO and  \ion{C}{2} emission), these authors found evidence of molecular outflows with a wide range of spatial extents (i.e., from few kpc up to 30 kpc) in the halo of QSOs at redshifts $4<\textit{z}<6$. 
In particular, a large fraction of the cold gas surrounding the quasars has been detected at higher velocities with respect to the systemic ones (average excess of 1000 km\,s$^{-1}$) and has been associated with possibly star-forming regions. 
Their most plausible interpretation of the outflow properties, including mass, momentum rate, mass-loss rate and kinetic power, is compatible with the predictions of the AGN-driven outflow models.
Unfortunately, to date, we have no information on possible CO and \ion{C}{2} extended emission from J0952+01144. Taken at face values, however, the molecular outflows velocities and spatial extent reported above are similar to the scenario presented in this discussion, which suggests that the possible outflow associated with the PDLA could be consistent with other molecular outflows detected with ALMA.

\subsection{J0952+0114 QSO: Exotic or ordinary object?}

If the scenario presented above is correct, then the QSO J0952+0114 is likely not a truly `exotic'  or extraordinary object but a QSO with similar intrinsic properties and environment with respect to other luminous QSOs at this redshift, but with the serendipitous presence of a PDLA absorber along our line-of-sight.
Indeed, the giant Ly$\alpha$ nebula that we have discovered around J0952+0114 is very similar to those observed by B16 and A19, which targeted typical QSOs without PDLAs. 
Other clumpy outflows can be present in other sources and the rarity of such findings could reflect the small covering factor of such outflows rather than the peculiarity of J0952+0114, though dedicated follow-up observations, similar to those presented here, would be needed to confirm it with certainty.

\vspace{1cm}
\section{Summary and Conclusions} \label{sec:conclusions}

In this paper, we presented the detailed study of 1 hour MUSE integral-field spectroscopic observations of the radio-quiet QSO J0952+0114 at $z$\,=\,3.02 with the goal of understanding the origin, location and spatially extent of the gas associated with the PDLA.
Our main findings can be summarized as follows:
\begin{itemize}
\item[] $\ast$ We found a giant ($\approx$ 100 pkpc) Ly$\alpha$ nebula at \textit{z}\,$\sim$\,3.017 around J0952+0114. The Ly$\alpha$ properties of the gas, i$.$ size, luminosity and SB profile, enshrouding this QSO are not different from other MUSE QSO giant nebulae without PDLAs.
\item[] $\ast$ We also detected the presence of narrow and extended \ion{C}{4} and \ion{He}{2} emission from the nebula at same redshift as Ly$\alpha$. 
\item[] $\ast$ In addition, we detected a bright spatially unresolved and relatively broad \ion{He}{2} emission in the central region at the redshift of the PDLA which possibly suggests the presence of a clumpy outflow on small scales that contains the PDLA.
\item[] $\ast$ We investigated the origin of the central regions of the nebula by analyzing the SB and velocity maps. The absence of abrupt changes in both properties suggests that we are witnessing a smooth transition between the CGM and the ISM in the region illuminated by the QSO. 
\item[] $\ast$ Moreover, we performed a kinematical analysis based on both the velocity and velocity dispersion maps of the narrow Ly$\alpha$,  \ion{C}{4} and  \ion{He}{2} emission lines. The relatively-low measured $\vec{v}$ and $\sigma$ values do not strongly support the idea of an ongoing outflow on large scales.
\vspace{0.2cm}
\item[] $\ast$ The analysis of the 2D \ion{He}{2}/Ly$\alpha$ and \ion{C}{4}/Ly$\alpha$ line ratio maps revealed a positive gradient with the distance to the QSO suggesting a non-homogeneous ionization parameter distribution of the CGM. 
\end{itemize}

Thanks to the MUSE's performance, we have shown that the unusual QSO J0952+0114 has the same extended Ly$\alpha$  nebula as essentially all the radio-quiet QSOs at this redshift. Because of the fortuitous presence of the PDLA, this particular system enables us to shed more light on the physical properties of the relatively cold gas around J0952+0114 with unprecedented details. \\


\acknowledgments{
 {\it Acknowledgments.} 
We thank Lutz Wisotzki for stimulating discussions.
This work is based on observations taken at ESO/VLT in Paranal and we would like to thank the ESO staff for their assistance and support during the MUSE GTO campaigns. This work was supported by the Swiss National Science Foundation. This research made use of \textit{Astropy}, a community$-$developed core PYTHON package for  astronomy \citep{2013A&A...558A..33A}, \textit{NumPy} and \textit{SciPy} (Oliphant 2007), \textit{Matplotlib} (Hunter 2007), \textit{IPython} (Perez \& Granger 2007), and of the NASA Astrophysics Data System Bibliographic Services. SC and GP gratefully acknowledge support from Swiss National Science Foundation grant PP00P2$_{-}$163824. AF acknowledges support from the ERC via Advanced Grant under grants agreement no. 339659-MUSICOS. JB acknowledges support by FCT/MCTES through national funds by grant UID/FIS/04434/2019 and through Investigador FCT Contract No. IF/01654/2014/CP1215/CT0003. SDJ is supported by a NASA Hubble Fellowship (HST-HF2-51375.001-A). TN acknowledges the Nederlandse Organisatie voor Wetenschappelijk Onderzoek (NWO) top grant TOP1.16.057.}

\newpage

\appendix

\setcounter{figure}{0} \renewcommand{\thefigure}{A.\arabic{figure}}
\setcounter{table}{0} \renewcommand{\thetable}{A.\arabic{table}}
\section{Identification of possible spectral features at the redshift of the PDLA.}

\begin{figure*}[ht]
\centering
\includegraphics[scale=0.32]{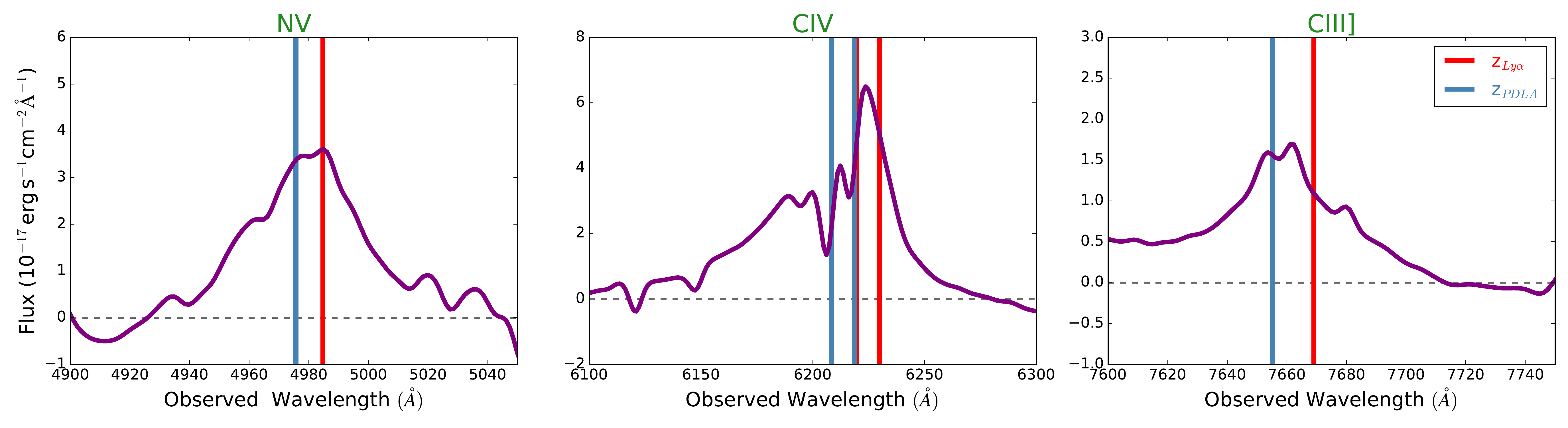}
\caption{Expanded view of the \ion{N}{5}, \ion{C}{4} and \ion{C}{3]} emission line regions. Blue vertical lines indicate the expected emission peaks corresponding to the PDLA redshift while the red ones correspond the redshift obtained from the narrow Ly$\alpha$ emission.\label{fig:PDLA_lines}}
\end{figure*}

In this Section we show the results from the analysis of the J0952+0114 MUSE spectrum focused on the identification of possible emission and absorption features at the PDLA redshift. We performed the line identification in the continuum subtracted spectrum of a 0.6$\arcsec$-diameter region centered on the QSO. The only clear isolated emission line, located at the PDLA redshift, is the \ion{He}{2} emission already presented in Fig.\,\ref{fig:HeII}. Fig.\,\ref{fig:PDLA_lines} shows the other features detected at PDLA redshift including a clear absorption system in the \ion{C}{4} wavelength region (middle panel) and \ion{N}{5} and \ion{C}{3]} emission lines (left and right panels, respectively).

\vspace{1cm}

\section{Comparison between the J0952+0114 MUSE spectrum and literature data.}

In this Section, we provide a comparison between the J0952+0114 MUSE spectrum and literature data. In order to perform a consistent comparison with the SDSS spectrum (green solid line), we extracted our J0952+0114 spectrum (blue solid line) from a 3$\arcsec$-diameter region. We note that the agreement between MUSE and SDSS is excellent. In addition, in Fig.\,\ref{fig:MUSE_SDSS_BOSS} also includes the comparison with the BOSS-QSO composite spectrum of \citealt{2016ApJ...833..199J} that comprises 3106 QSO spectra with a median bolometric luminosity log(L$_{\mathrm bol}$)\,=\,46.66, a median spectral index  $\alpha$\,= -1.52 and a median redshift \textit{z} of 2.91. This particular composite spectrum has been chosen because it better matches the J0952+0114 properties (i$.$e$.$ log(L$_{\mathrm bol}$)\,=\,46.48 and \textit{z}=3.02). By comparing our J0952+0114 MUSE spectrum with the BOSS-QSO composite one, we found that the position and shape of the J0952+0114 \ion{He}{2} emission line are significantly different from the one of other QSOs with similar luminosity, redshift, and UV slope as highlighted in Fig.\,\ref{fig:MUSE_SDSS_BOSS_zoom}.

\setcounter{figure}{0} \renewcommand{\thefigure}{B.\arabic{figure}}
\begin{figure}[ht]
\centering
\includegraphics[scale=0.2]{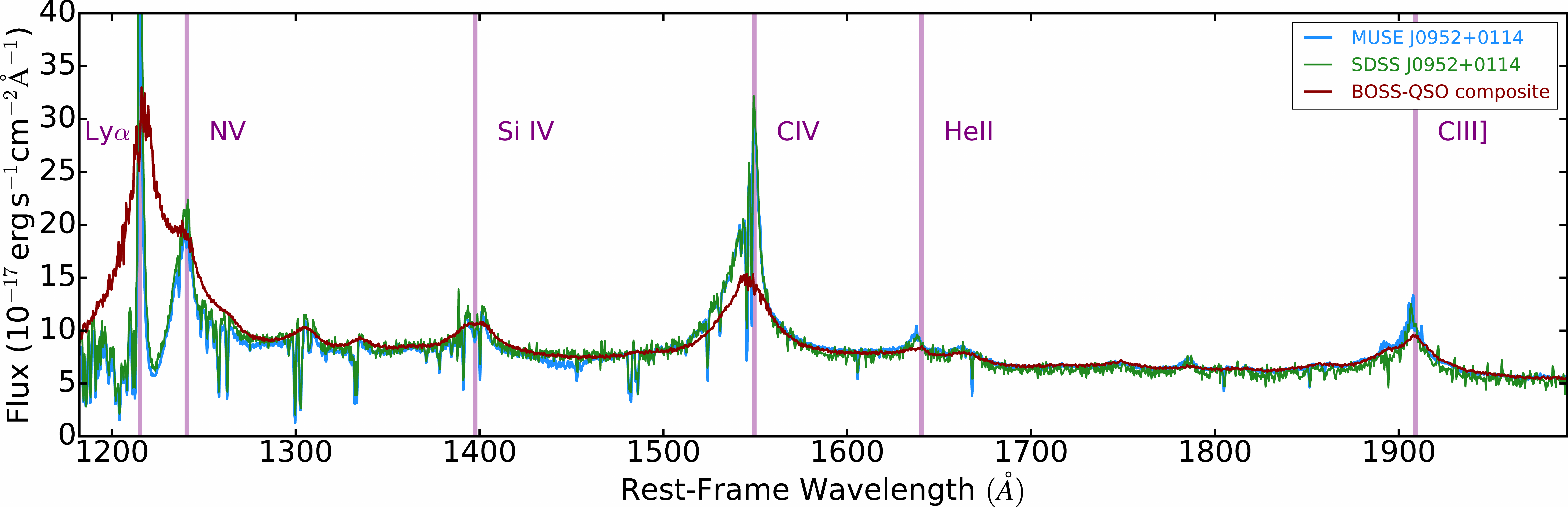}
\caption{ Comparison between the MUSE (blue), the SDSS (green), and BOSS-QSO composite (dark red) spectra.  \label{fig:MUSE_SDSS_BOSS}}
\end{figure}

\begin{figure*}[h!]
\centering
\includegraphics[width=0.45\columnwidth]{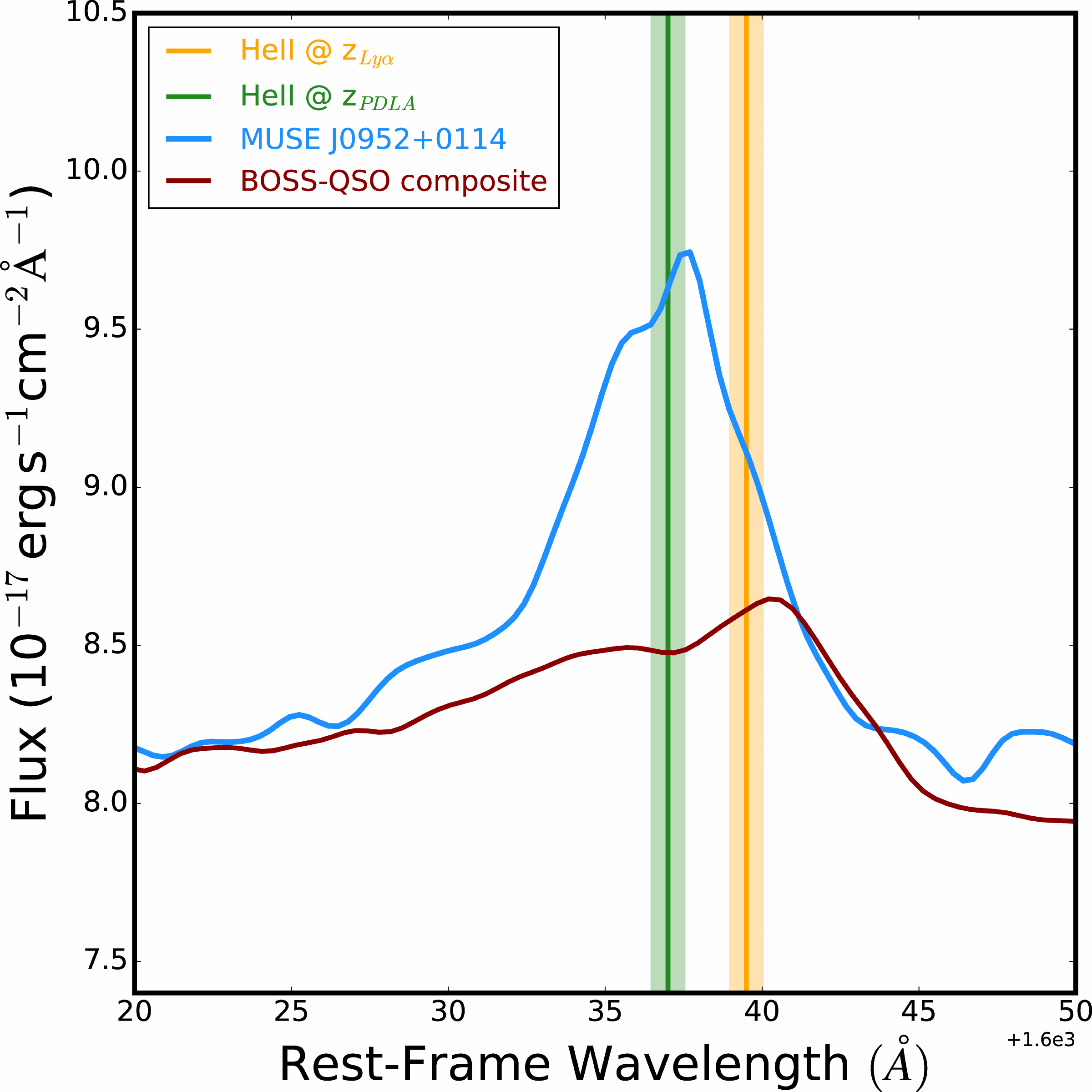}
\caption{Zoom-in on the \ion{He}{2} emission lines observed in the MUSE (blue) and the BOSS-QSO composite (dark red) spectra. The green and the orange vertical lines/areas indicate the position of the \ion{He}{2} emission in case of adopting the redshifts of the PDLA and the nebula, respectively.\\ \label{fig:MUSE_SDSS_BOSS_zoom}}
\end{figure*}

\newpage

\section{Comparison between the J0952+0114 and literature Ly$\alpha$ SB profiles.}

As shown in Fig.\,\ref{fig:SB}, we have compared the circularly averaged SB profile of the J0952+0114  Ly$\alpha$ nebula with other Ly$\alpha$ SB profiles already published in the literature, i$.$ e$.$ B16 and A19. In the case of A19 we plotted an exponential profile (blue solid line) using the average values presented in their Table 4. In the case of the B16 MUSE Ly$\alpha$ SB profiles, we plotted the median profile (solid purple line) together with the 10th and 90th percentiles that are given in Table\,\ref{tab:B16values}.

\setcounter{table}{0} \renewcommand{\thetable}{C.\arabic{table}}
\begin{table}[h!]
\centering
\caption{Avarage, median and percentiles values of the MUSE Ly$\alpha$ SB profiles published in \citealt{2016ApJ...831...39B}}  \label{tab:B16values}
\begin{tabular}{c c c c c c c}
\tablewidth{0pt}
\hline
\hline
Radius & average  Ly$\alpha$ SB & median  Ly$\alpha$ SB & 10th percentile & 25th percentile & 75th percentile & 90th percentile \\
\hline
[kpc] &  \multicolumn6c{[10$^{-18}$ erg s$^{-1}$ cm$^{-2}$ arcsec$^{-2}$]}\\
\hline
\hline
   8.7     &   58.38 &  35.24  &  14.05  &  28.76   & 63.08   & 143.06   \\
   10.9    &   45.78 &  30.92  &  8.19   &  15.76   & 44.78   & 116.99   \\
   13.7    &   32.45 &  25.79  &  6.17   &  9.18    & 28.58   & 82.33   \\
   17.3    &   23.80 &  15.07  &  4.81   &  5.50    & 18.22   & 57.84   \\
   21.7    &   16.79 &  10.59  &  4.44   &  6.01    & 13.26   & 37.20   \\
   27.4    &   11.80 &  6.67   &  2.02   &  3.83    & 10.73   & 29.00   \\
   34.5    &   7.88  &  5.89   &  1.37   &  3.35    & 6.92    & 17.59   \\
   43.4    &   5.56  &  4.29   &  1.06   &  2.39    & 6.10    & 11.30   \\
   54.6    &   2.23  &  1.89   &  0.42   &  1.41    & 2.68    & 3.49   \\
   68.8    &   0.97  &  0.64   &  0.05   &  0.27    & 1.52    & 2.00   \\
   86.6    &   0.30  &  0.23   & -0.48   & -0.11    & 0.41    & 1.25   \\
   109.0   &   0.07  &  0.10   & -0.71   & -0.05    & 0.21    & 0.55 \\

\hline
\multicolumn{7}{c}{NOTE. - All SB values were rescaled to  \textit{z}\,=\,3.}
\end{tabular}
\end{table}



\end{document}